\documentclass[%
 reprint,
 amsmath,amssymb,
 aps,
xcolor,
hyperref,
floatfix,
]{revtex4-2}
\usepackage{graphicx}% Include figure files
\usepackage{dcolumn}% Align table columns on decimal point
\usepackage{bm}% bold math
\DeclareMathOperator{\EX}{\mathbb{E}}
\DeclareMathOperator{\VAR}{\mathbb{V}ar}
\begin{document}

\preprint{APS/123-QED}
%\title{Eigenvalue spectral properties of sparse random matrices for neural networks obeying Dale's law.}
%\title{Consequences of sparse connectivity on random neural networks obeying Dale's law}

\title{The effect of sparsity on network stability in random neural networks obeying Dale's law}% Force line breaks with \\

\author{Isabelle D. Harris}
\email{harrisi@unimelb.edu.au}
\affiliation{Department of Biomedical Engineering, University of Melbourne, Australia\\
 Graeme Clark Institute, University of Melbourne, Australia
}%


\author{Hamish Meffin}
\affiliation{Department of Biomedical Engineering, University of Melbourne, Australia\\
National Vision Research Institute, Australian College of Optometry, Melbourne, Australia
}%

\author{Anthony N. Burkitt}
\affiliation{Department of Biomedical Engineering, University of Melbourne, Australia\\
 Graeme Clark Institute, University of Melbourne, Melbourne, Australia
}%

\author{Andre D. H. Peterson}
\affiliation{Department of Biomedical Engineering, University of Melbourne, Australia\\
 Graeme Clark Institute, University of Melbourne, Australia\\
 Department of Medicine, St. Vincent’s Hospital, University of Melbourne, Australia
}%


\date{\today}% It is always \today, today,

\begin{abstract}
This paper examines the relationship between sparse random network architectures and neural network stability by examining the eigenvalue spectral distribution. Specifically, we generalise classical eigenspectral results to sparse (not fully-connected) connectivity matrices obeying Dale’s law: neurons function as either excitatory (E) or inhibitory (I). By defining $\alpha$ as the probability that a neuron is connected to another neuron, we give explicit formulae that shows how sparsity interacts with the E/I population statistics to scale key features of the eigenspectrum, in both the balanced and unbalanced cases. Our results show that the eigenspectral outlier is linearly scaled by $\alpha$, but the eigenspectral radius and density now depends on a nonlinear interaction between $\alpha$ and the E/I population means and variances. Contrary to previous results, we demonstrate that a non-uniform eigenspectral density results if any of the E/I population statistics differ, not just the variances. We also find that ‘local’ eigenvalue-outliers are present for sparse random matrices obeying Dale’s law, and demonstrate that these eigenvalues can be controlled by a modified zero row-sum constraint for the balanced case, however, they persist in the unbalanced case. We examine all levels of connection sparsity $0 \leq \alpha \leq 1$, and distributed E/I population weights, to describe a general class of sparse connectivity structures which unifies all the previous results as special cases of our framework. Sparsity and Dale’s law are both fundamental anatomical properties of biological neural networks. We generalise their combined effects on the eigenspectrum of random neural networks, thereby gaining insight into network stability, state transitions and the structure-function relationship.

\end{abstract}

\maketitle

%\tableofcontents

\section{\label{sec:introduction}Introduction\protect}
     % [intro]
    Understanding the spatio-temporal dynamics of large populations of neurons in the cortex is a fundamentally difficult open problem in both theoretical and experimental neuroscience, particularly the relationship between network connectivity and dynamics. Theoretically, this has been typically studied by either averaging over the synaptic connection weights, thereby sacrificing network structure, or via large-scale numerical simulations of neural models that are mathematically intractable. An effective approach that preserves the statistical structure of the synaptic connectivity whilst still being mathematical feasible is to study the dynamics of partially random networks of neurons. Network dynamics in this framework are examined through changes in the eigenvalue spectral distribution of the network Jacobian, which is a function of the synaptic connectivity matrix \cite{rajan2006eigenvalue}. This paper mathematically examines the stability properties of the Jacobian's eigenspectrum when more realistic anatomical structure is incorporated into the connectivity matrix, such as sparsity, network (im)balance, and Dale's law. \\
    % [sparsity]
    A key feature of biological neural networks is that they are not fully-connected, namely neurons do not receive input from every other neuron in the network \cite{brunel2000dynamics,barral2016synaptic,golomb2000number}. Usually the number of connections is relatively small, but varies depending on spatial scale, location, network size, and specific population wiring related to function. We introduce sparsity into the synaptic connectivity matrix by defining a sparsity parameter, $\alpha$, as the probability that a neuron is connected to another neuron, so that $\alpha$=1, denotes a fully connected network. Previous analyses only considered fully connected networks \cite{rajan2006eigenvalue,stern2014dynamics,landau2018coherent,ipsen2020consequences}, sparse networks with constant weights describing each of the excitatory and inhibitory populations \cite{ostojic2014two,mastrogiuseppe2017intrinsically, herbert2022impact},  or sparse one population random networks \cite{tao2008random,herbert2022impact}. Furthermore, these works \cite{ostojic2014two,mastrogiuseppe2017intrinsically} are only valid in the very sparse limit, i.e., $\alpha<<1$. This paper generalises the previous results to include all levels of sparsity $0\leq \alpha\leq1$ in two-population networks with differently distributed weights and different network (im)balances. \\
    % [situating research problem in current balance literature]
    There is a considerable amount of experimental \cite{barral2016synaptic,marino2005invariant,gorur2022mapping} and theoretical \cite{van1996chaos,brunel2000dynamics,brunel2003determines,staley2015molecular} evidence that strongly suggests that brain activity crucially depends on the dynamic balance between excitation and inhibition, and is essential for brain function \cite{barral2016synaptic,landau2018coherent}. Many anatomical and physiological network properties adjust homeostatically to maintain balanced E-I input \cite{marder2006variability}, and network imbalances can lead to pathological brain dynamics, such as epileptic seizures \cite{ipsen2020consequences}. However, the concept of network balance is ambiguous \cite{ahmadian2021dynamical} and needs to be defined carefully. Functional network balance is a dynamical property that changes depending on the network activity. Specifically, we define functional network balance as the sum of synaptic inputs, i.e., the weights multiplied by the firing rates \cite{meffin2004analytical,brunel2000dynamics}. However, in this work we do not consider the firing rates, and focus instead on structural network balance. Structural network balance in biological neural networks is intrinsically tied to Dale's law, where neurons in the cortex are either excitatory (E) or inhibitory (I) in their action on target neurons \cite{eccles1976electrical}. Dale's law introduces a macroscopic anatomical constraint upon the random synaptic connectivity matrix; i.e. a partially random neural network. Therefore, we define structural E-I network balance to be the network state in which the mean excitatory (E) weights equals the mean magnitude of the inhibitory (I) weights \cite{barral2016synaptic,brunel2000dynamics,meffin2004analytical}. We examine both structurally balanced and unbalanced networks in combination with sparsity in this framework to understand their impacts on brain dynamics, particularly state transitions to physiologically realistic asynchronous activity \cite{ostojic2014two}. \\
    In this paper, we consider the combined effects on network stability of incorporating both sparsity and Dale's law.  %In particular, we pose the question: how does incorporating both sparsity and Dale's law into the connectivity matrix affect the stability and density of the eigenspectrum? 
     We commence by reviewing previous results related to random neural networks (Section~\ref{sec:modelandanalysis}), and eigenvalue spectral properties of synaptic connectivity matrices (Section~\ref{sec:evalsrandommatrix}-Section~\ref{sec:sparserandommatrix}). In Section~\ref{sec:evalssparserandommatrix} we extend these previous results by analysing the eigenvalue distribution (including outliers)  of the networked Jacobian for sparse balanced and unbalanced random synaptic connectivity matrices obeying Dale's law.  Specifically, we deduce a number of mathematically explicit formulas that extend previous analyses \cite{rajan2006eigenvalue,ipsen2020consequences,herbert2022impact}, yielding a quantitative relationship between sparsity, the E-I populations statistics and principal properties of the eigenspectrum.  

\section{\label{sec:modelandanalysis}Network model and analysis\protect} 
    In this study the neural network dynamics are described by
    \begin{equation}
        \dot{x}_{i}(t) = -\frac{x_{i}(t)}{\tau} + \sum_{j=1}^{N} w_{ij} \phi(x_{j}(t)), \label{Equ:Model}
        \end{equation}
    where $x_{i}(t)$ is the current of the $i$th neural unit, $\tau$ is the time constant, $w_{ij}$ is an entry in a $N \times N $ partially random network connectivity matrix $W$, $\phi(x_{i}(t))$ is an activity-to-firing rate coupling function. The function $\phi$ is defined as a real valued, bounded, smooth, and strictly monotonically increasing odd function on the infinite domain with $\phi(0)=0$, $\phi'(0)=1$, and $\phi \rightarrow \pm 1$ for $x \rightarrow \pm \infty$, e.g., $\phi(x) = \tanh(x)$ \cite{sompolinsky1988chaos,stern2014dynamics,ostojic2014two,ipsen2020consequences}. %For such a function, the system yields both positive and negative outputs. This can be interpreted as a firing rate where $\phi$ is the deviation of the output away from a homogeneous average or baseline rate for the network, which can be considered as the homeostatic fixed point of the network dynamics, and $\phi(-\infty)$ is the zero-point of the firing rate scale.
        
    %\textbf{[system analysis]} \par
    The equilibria of this network model are the solutions of the general expression
    \begin{equation}
        \boldsymbol{x^*} = \tau W \boldsymbol{\phi(x^*)}, \label{equ:equilibria}
    \end{equation}
    where $\boldsymbol{x^*}, \boldsymbol{\phi(x^*)}  \in \mathbb{R}^N$. Hence, these solutions are directly dependent on the structure of the connectivity matrix $W$. Networks described by Eq.~\ref{Equ:Model} and a random connectivity matrix with zero mean always yield a `trivial' homogeneous equilibrium solution. However, in the case of random networks obeying Dale's law the existence of a homogeneous equilibrium solution requires that the sum of the rows of the connectivity matrix is equal across all rows, formally,
    \begin{equation}
        \sum_{j}^{N} w_{ij} = N \mu_{\mathrm{r}} \label{equ:rowsumW},
    \end{equation}
    where $\mu_r$ is the average connectivity weight. If Eq.~\ref{equ:rowsumW} is satisfied, then there exists a homogeneous equilibrium solution when $x_{k}^{*} = x_{0}^{*}$ provided that 
    \begin{equation}
        x_{0}^{*} = \tau N \mu_{\mathrm{r}} \phi(x_{0}^{*}), \label{equ:FP}
    \end{equation}
    has a solution for all units $k=1,...,N$. 
    %Calculating the equilibrium solutions of Eq.~\ref{equ:equilibria} for large neural networks is both mathematically intractable and computationally infeasible, since the number of equilibria scales exponentially with system size \cite{ipsen2020consequences,wainrib2013topological}.
    %Structural (E-I) balance or imbalance is defined as the  weighted contributions of the average neural population strengths. Therefore the expected value of the entries of the rows of $W$, namely the value of $\mu_{\mathrm{r}}$, indicates whether the network is balanced. If the connectivity matrix is excitatory dominated then $\mu_{\mathrm{r}}>0$, or inhibitory dominated $\mu_{\mathrm{r}}<0$, and it is structurally E-I balanced when $\mu_{\mathrm{r}}=0$. 
    Networks that satisfy the row-sum condition and structural (E-I) balance, yield a `trivial' homogeneous equilibrium solution, $\boldsymbol{x_{0}^{*}}=\boldsymbol{0}$. If however, a network satisfies the row-sum condition, but is structurally E-I unbalanced, a constant homogeneous equilibrium solution exists, $\boldsymbol{x_{0}^{*}}=\boldsymbol{\xi}$. Alternatively, if the row-sum condition is not satisfied and the network is structurally E-I unbalanced, then the system permits heterogeneous equilibria, i.e., different neurons $i$ attain two or more different equilibrium values.%Heterogeneous equilibrium solutions $\boldsymbol{x^*}$ are difficult to calculate, and are dependent on the individual realisation of the connectivity matrix \cite{mastrogiuseppe2018linking,harish2015asynchronous}.
    
    To evaluate the local stability of the system around the equilibria, we study the eigenspectrum of the networked Jacobian 
    \begin{equation}
        \mathcal{J}(\boldsymbol{x}^{*}) = \left[-\frac{1}{\tau} \mathbb{I}_N + W \Phi'(\boldsymbol{x^*})\right], \label{Equ:Jacobian}
    \end{equation}
    where, $\mathbb{I}_N$ is the identity matrix, and $\Phi'(\boldsymbol{x^*})$ is a $N\times N$ matrix.
    % diagonal matrix with diagonal entries 
    % \begin{equation}
    %     \frac{\partial \phi(x_{j})}{\partial x_{j}}\vert_{\boldsymbol{x^*}} = \phi' (x^*_j).
    % \end{equation} 
    When the real part of at least one of the eigenvalues of the Jacobian Eq.~\ref{Equ:Jacobian}, becomes positive, the equilibrium solution becomes unstable and spontaneous dynamics emerge \cite{allesina2015stability,mastrogiuseppe2017intrinsically}. Hence, the local stability and neural dynamics is influenced by the eigenspectrum of the Jacobian, which from random matrix theory, depends on the statistical structure of the synaptic connectivity matrix~$W$.
    %Note that after destabilisation of the network, this analysis is no longer useful the transition has occurred to 
    
     If the homogeneous equilibrium is the `trivial' zero solution,  $\boldsymbol{x_{0}^{*}}=\boldsymbol{0}$, then the matrix $\Phi'(\boldsymbol{x_{0}^{*}}) = \mathbb{I}_N$ since $\phi'(0) = 1$. Thus, the eigenspectrum of the Jacobian depends only on the synaptic connectivity matrix $W$ with diagonal offset of $-1/\tau$. However, if the homogeneous equilibria is a constant value, $\xi$ for all units $i$, $\boldsymbol{x_{0}^{*}}=\boldsymbol{\xi}$, the matrix $\Phi'(\boldsymbol{x^*}) = \gamma \mathbb{I}_N$ since $\phi' (\boldsymbol{\xi}) = \boldsymbol{\gamma}$. This introduces an additional dependence in the Jacobian without changing the overall statistical structure, as $\gamma$ only scales all of the connectivity strengths of $W$. 

    In contrast, the network Jacobian of heterogeneous equilibrium solutions, $\boldsymbol{x^*}$, incorporates an additional (nested) dependence of the connectivity matrix through the term $\Phi'(\boldsymbol{x^{*}})$. $\Phi'(\boldsymbol{x^{*}})$ correlates the structured and random components of the connectivity matrix. Further, this term correlates the Jacobian, $\mathcal{J}(\boldsymbol{x}^{*})$, to the individual realisation of the random part of the connectivity matrix \cite{mastrogiuseppe2018linking}.  Therefore, to examine the influence of the statistical structure of the connectivity matrix $W$ on heterogeneous equilibrium solutions, $\boldsymbol{x^*}$, the eigenvalues of the Jacobian, and hence the network dynamics, dynamical mean-field techniques are required \cite{mastrogiuseppe2018linking,harish2015asynchronous}. This analysis is outside the scope of this investigation. 

    In this paper, we focus our analysis on the effects of implementing anatomically realistic structure into a random connectivity matrix, namely Dale's law, structural E-I imbalance, and sparsity, on the eigenvalue spectral distribution of Eq.~\ref{Equ:Jacobian} evaluated at $\boldsymbol{x_{0}^{*}}=\boldsymbol{0}$ and assume a unit scaling factor $\phi'(0) = 1$.   

\section{\label{sec:eigenspectralproperties}Eigenvalue spectral properties of synaptic connectivity matrices} 
    To investigate the impact of Dale's law, structural E-I imbalance, and sparsity in random neural networks obeying Dale's law, we analyse the changes in the eigenspectral distribution of the network Jacobian Eq.~\ref{Equ:Jacobian}. We use the following key result from random matrix theory to examine this relationship. 
    \subsection{\label{sec:evalsrandommatrix} Eigenvalues of a random matrix.}
        The elements in large synaptic connectivity matrices are sampled randomly from a Gaussian or any identically independent distribution. Therefore, we use results from random matrix theory in this investigation \cite{rajan2006eigenvalue,ipsen2020consequences}. A central result of random matrix theory is Girko's circular law. This law states that the empirical spectral distribution of a random matrix, $A$, with entries $a_{ij}$ independently and identically distributed (iid) with mean $\mu=0$, variance $\sigma^2 = \frac{1}{N}$, converges to the unit disc on the complex plane \cite{mehta2004random,girko1985circular,rajan2006eigenvalue,tao2008random, tao2010random}. A secondary result states that an eigenvalue outlier escapes the eigenspectral-disc if $A$ has nonzero mean \cite{tao2013outliers}. Consequently an iid random matrix $A_N$ with mean $\mu \neq 0$, variance $\sigma^2 = \frac{1}{N}$, and finite fourth moment has an eigenspectrum with a central eigenspectral-disc, with radius $\mathcal{R}=\sigma \sqrt{N}$, and a single eigenvalue outlier, $\lambda_O$, that escapes to the point $\lambda_O = \mu N$ on the complex plane. 
    
    \subsection{\label{sec:sparserandommatrix}Sparse random matrices}
         Previous work studying sparse random matrices examined the eigenvalue spectral distribution of Boolean random matrices, sparsified Gaussian random matrices \cite{tao2008random,herbert2022impact}, sparsified low-rank networks \cite{herbert2022impact}. Two previous papers of note, \cite{mastrogiuseppe2017intrinsically,shao2023relating} study sparse random matrices obeying Dale's law, however Dale's law is implemented by setting all excitatory and inhibitory units to constants, $\bar{W}_e$, $\bar{W}_i$, essentially removing the underlying random distribution of connectivity elements before sparsity is applied. In this paper, we examine a more general class of sparse random matrices and we note that these previously studied matrices and corresponding results are all special cases of the following class of sparse random matrices. 
               
        %Review paragraph: Herbert - Eigenvalues of sparsified full-rank networks (sparse AD) and Eigenvalues of sparsified rank-one networks (sparse M where the elements of the outer product of M=\boldsymbol{uv}T are distributed randomly). Shao - sparsified Networks obeying Dale's law (E-I) implemented as constants. These previous results all form special cases of the more general results we present in the next section .
                
        A sparse random matrix, $W$, is characterised by three statistics; the probability of a nonzero element $\alpha$, the mean $\mu$, and variance $\sigma^{2}$ of the nonzero entries. We construct our connectivity matrix $W$ by combining sparse, random and deterministic components as per Eq.(\ref{equ:sWrankperturb}) below. $A D$ is the random component, and $M$ is a low-rank deterministic component. Incorporating network sparsity is achieved by a Hadamard (element-wise) product of $A D + M$ with a Boolean random matrix $S$. A sparse random matrix is defined as,
        \begin{equation}
            W = S \circ (A D + M) \label{equ:sWrankperturb},
        \end{equation}
        where $S$ is an iid Boolean random matrix with probability $\alpha$ of an element being non-zero, $\circ$ is the element wise product, $A$ is an iid random matrix with zero mean and unit variance, $D$ is a diagonal matrix of standard deviations, $D = \text{diag}(\sigma, ...., \sigma)$, $M = \boldsymbol{uv}^\top$ is a rank one matrix perturbation with row vectors $\boldsymbol{u} = (1, ..., 1)^\top$, $\boldsymbol{v} = (\mu, ..., \mu)^\top$. If $\mu \neq 0$ then the connectivity matrix $W$ is structurally (E-I) unbalanced.
        
        We scale the mean and standard deviation, by $\sqrt{N}$ to ensure that the properties of the eigenvalue spectrum are as independent as possible from the system size. Specifically, we use scaled variables $\tilde{\mu} = \frac{\mu}{\sqrt{N}}$, and $\tilde{\sigma} = \frac{\sigma}{\sqrt{N}}$. 

    \subsection{\label{sec:evalssparserandommatrix}Eigenvalues of sparse random matrices}
        \begin{figure}
         \centering
                %\subfloat[]{%
                    \includegraphics[clip,width=1\columnwidth]{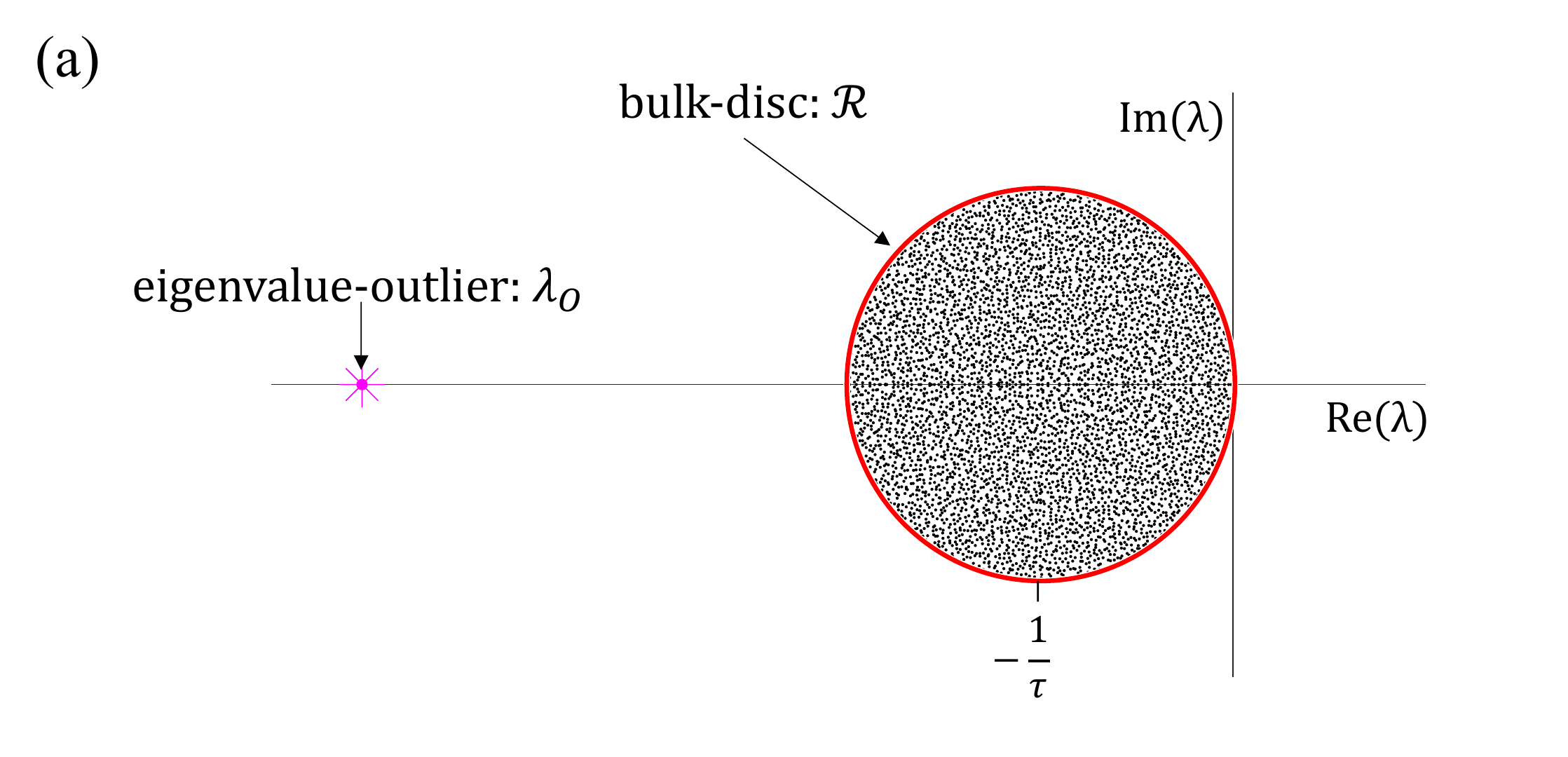}%
                %}
                \quad
                %\subfloat[]{%
                    \includegraphics[clip,width=1\columnwidth]{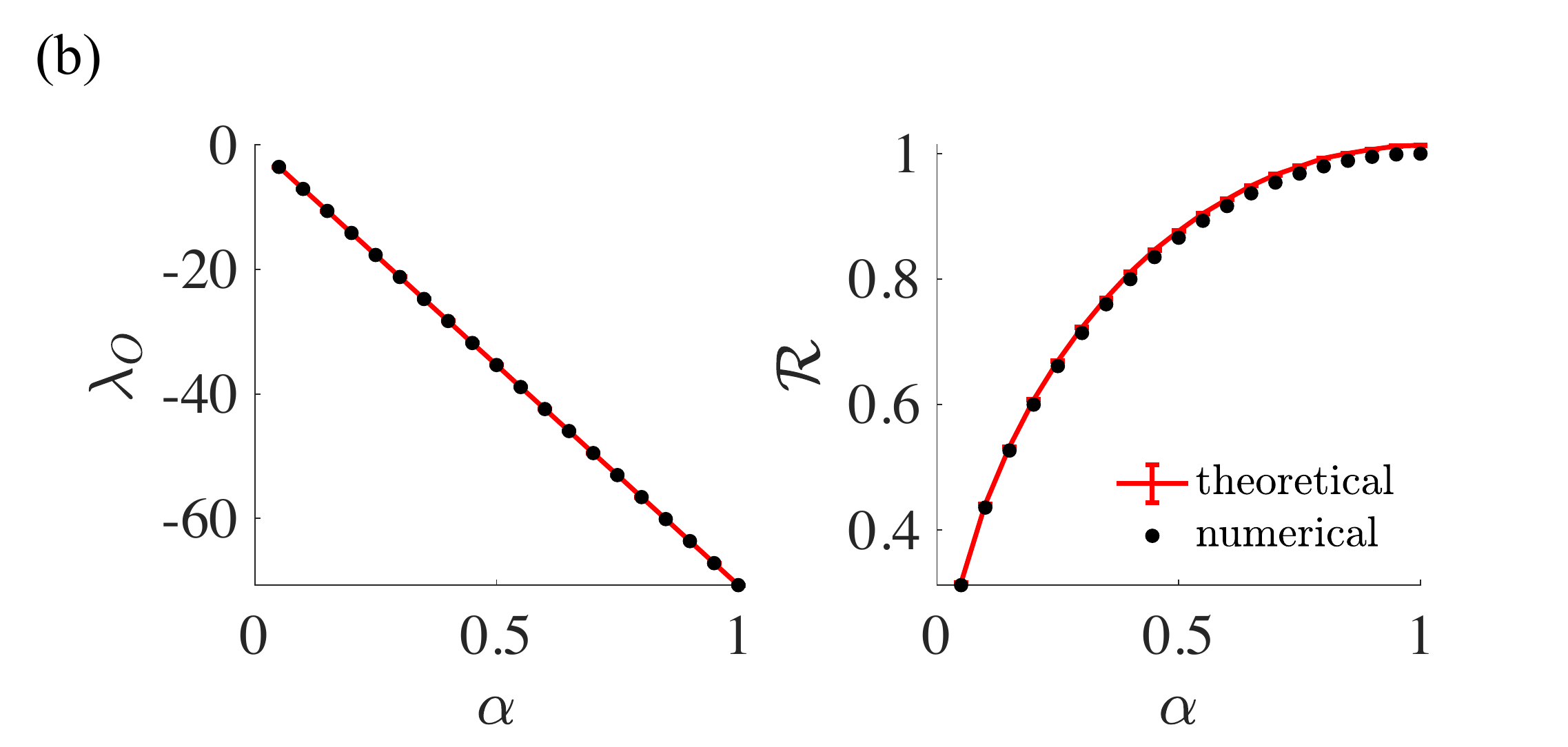}%
                %}
            \caption{(a) Eigenspectrum of $W$ (Eq.~\ref{equ:sWrankperturb}) with $N = 5000, \tilde{\mu} = -\frac{1}{\sqrt{N}},\tilde{\sigma} = \frac{1}{\sqrt{N}}, \alpha = 0.99$. (b) The eigenvalue outlier (left) and the radius of the eigenspectral-disc (right), plotted against the sparsity parameter $\alpha$, for the matrix $W$. The eigenvalue outlier $\lambda_o$ was calculated theoretically (black) from Eq.~\ref{equ:lambdao} and numerically (red) from $W$. The eigenspectral radius (black) was calculated theoretically using Eq.~\ref{equ:rad1b} and numerically by the average second largest eigenvalue of $W$. The numerical quantities in red were averaged over 100 realisations of $W$ and are shown with standard error bars.} \label{fig:onepopfigOR}
        \end{figure}
        A typical eigenspectrum for an structurally E-I imbalanced matrix $W$ is illustrated in Figure~\ref{fig:onepopfigOR}(a). The two primary properties are the location of the eigenvalue outlier, $\lambda_O$, and the radius of the eigenspectral-disc, $\mathcal{R}$. Results from random matrix theory predict that $\lambda_O = \EX(w_{ij}) N$, and $\mathcal{R} = \sqrt{\VAR(w_{ij})N}$ \cite{tao2008random,tao2010random,tao2013outliers,rajan2006eigenvalue,mehta2004random}. However, these results are only explicitly proven for fully connected networks with nonzero mean, and sparse random matrices with zero mean. 
        
        We build on these previous results to predict the location of $\lambda_O = \EX(w_{ij}) N$, and the radius $\mathcal{R} = \sqrt{\VAR(w_{ij})N}$ for sparse unbalanced random connectivity matrices. For sparse random matrices that obey Dale's law, we compute $\lambda_O$, and $\mathcal{R}$, by first deriving expressions for the means and variances of the weights $w_{ij}$. 
    
        We relate (Appendix~\ref{app:derivation1}) the mean of the weights, $\EX(w_{ij})$, to the normalised mean $\tilde{\mu}$, and sparsity parameter $\alpha$, by
        \begin{equation}
            \EX(w_{ij}) = \alpha \tilde{\mu} \label{equ:meaneff1}.
        \end{equation}
        Therefore, we predict that the location of the eigenvalue outlier is given by
        \begin{equation}
            \lambda_O = \alpha \tilde{\mu} N \label{equ:lambdao}.
        \end{equation}
        We compare the predicted eigenvalue outlier defined in Eq.~\ref{equ:lambdao} to the eigenvalue outlier in the numerical eigenspectrum of sparse random matrices constructed using Eq.~\ref{equ:sWrankperturb}. Numerical eigenspectra are calculated using MATLAB, and this is performed for a large number of realisations of the sparse random matrix. We compute the eigenvalue with the largest magnitude for each realisation, and average the eigenvalue-outliers over all realisations to obtain a numerical estimate for $\lambda_O$. The predicted eigenvalue outlier and average numerical eigenvalue outlier over 100 realisations of the matrix $W$ is shown in Figure~\ref{fig:onepopfigOR}(b).

        To calculate the radius of the eigenspectral disc, the variance of the weights, $\VAR(w_{ij}) = \EX(w_{ij}^{2}) - \EX(w_{ij})^2$ is derived as a function of the three primary statistics; the normalised mean $\tilde{\mu}$, the normalised standard deviation $\tilde{\sigma}$, and sparsity parameter $\alpha$, see Appendix~\ref{app:derivation1} for details. The expression for the variance is now dependent on both the mean and sparsity parameters,
        \begin{equation}
            \VAR(w_{ij}) = \alpha (1-\alpha) \tilde{\mu}^2  + \alpha \tilde{\sigma}^{2}  \label{equ:sigeff1}.
        \end{equation}
        Therefore, we can now compute the radius of the eigenspectral-disc as
        \begin{equation}
            \mathcal{R} = \sqrt{N\left[\alpha (1-\alpha) \tilde{\mu}^2 + \alpha \tilde{\sigma}^{2}\right]}. \label{equ:rad1b}
        \end{equation}
        Note that in accordance with the circular law \cite{tao2008random,tao2013outliers}, all the eigenvalues will converge to lie within a disc of radius $\mathcal{R}$ as $N \rightarrow \infty$. The expressions in Eq.~\ref{equ:lambdao} and Eq.~\ref{equ:rad1b} show that the normalised scaling ensures that the eigenvalue outlier location is of order $O(\sqrt{N})$ and the radius of order $O(1)$ as $N$ gets large. From this point, our analysis implicitly assumes this respective scaling.

        This choice of scaling is justified by our focus on balanced or inhibitory dominated networks that are close to balanced as opposed to excitatory dominated. In excitatory dominated networks, the eigenvalue outlier lies to the right of the disc and causes network activity to diverge and saturate to the upper bound of the firing rate function as the system size increases ($O(\sqrt{N})$). This activity is not of interest from a physiological perspective as it does not generate the spontaneous asynchronous activity associated with normal brain function, for example, the resting state \cite{ostojic2014two,mastrogiuseppe2017intrinsically,landau2018coherent}. Non-trivial spontaneous behaviour only emerges when the eigenspectral-disc crosses the imaginary axis and the network activity becomes unstable but not divergent. For this to occur, the network must be either balanced or inhibitory dominant so that the radius of the eigenspectral disc can grow with the variance of the connectivity matrix.
        
        Previous work on sparse random matrices with zero mean \cite{tao2008random} finds that the radius of the eigenspectral-disc scales linearly with the sparsity and the variance. By comparison, when the mean is non-zero, we find the radius is dependent on all three statistics $\tilde{\mu}$, $\tilde{\sigma}$, $\alpha$ and the system size $N$. Figure~\ref{fig:onepopfigOR} shows the predicted expression in Eq.~\ref{equ:rad1b} and the numerically calculated average radius of the eigenspectral-disc. To calculate the average radius, we extract the eigenvalue with the second largest magnitude for each realisation, and then average these values over all realisations. Based on results from random matrix theory, we know that the eigenvalue with the second largest magnitude should lie exactly on, or just within the radius of the eigenspectral disc \cite{tao2008random,tao2008random}. We find that there is agreement between the predicted eigenvalue outlier and radius and the numerical estimates of the eigenvalue outlier and radius, and the relative error between the estimates is of the order $10^{-4}$.

            \subsection{Eigenspectral properties of sparse random matrices that obey Dale's law} \label{sec:sparserandomDaleLaw}
        Distinct neural populations (Dale's law) are incorporated into the synaptic connectivity matrix by specifying two separate but related Gaussian distributions for each of the excitatory and inhibitory populations.  The sparse random matrix is partitioned into $Nf$ excitatory columns  ($\tilde{\mu}_{\mathrm{e}}$, $\tilde{\sigma}_{\mathrm{e}}^2$) and $N(1-f)$ inhibitory columns ($\tilde{\mu}_{\mathrm{i}}$, $\tilde{\sigma}_{\mathrm{i}}^2$) . The synaptic connectivity matrix still takes the form of Eq.(\ref{equ:sWrankperturb}), where $S$, and $A$ are defined as before. However, $D$ is now a diagonal matrix of excitatory and inhibitory variances, 
        \begin{equation}
            D=\text{diag}(\underbrace{\tilde{\sigma}_{\mathrm{e}},\ldots,\tilde{\sigma}_{\mathrm{e}}}_{Nf\text{ times}},\underbrace{\tilde{\sigma}_{\mathrm{i}},\ldots,\tilde{\sigma}_{\mathrm{i}}}_{N(1-f)\text{ times}}), \label{equ:Sigma}
        \end{equation}
        and the perturbation $M = \boldsymbol{uv}^\top$ is an outerproduct matrix of population means, with 
        \begin{equation} \label{equ:uv}
            u=(1,\ldots,1)^\top,\quad\ 
            v=(\underbrace{\tilde{\mu}_{\mathrm{e}},\ldots,\tilde{\mu}_{\mathrm{e}}}_{Nf\text{ times}},\underbrace{\tilde{\mu}_{\mathrm{i}},\ldots,\tilde{\mu}_{\mathrm{i}}}_{N(1-f)\text{ times}})^\top.
        \end{equation}
        The matrix $M$ consists of $Nf$ columns with identical entries  $\tilde{\mu}_{\mathrm{e}}$ and the remaining $N(1-f)$ columns with entries $\tilde{\mu}_{\mathrm{i}}$. Here, structural (E-I) balance is defined in terms of the combined relative contributions of the excitatory and inhibitory neurons, i.e., the expected value of the entries, $\EX(w_{ij})$. We exploit this formalism to examine sparse structurally (E-I) unbalanced synaptic connectivity matrices, $\EX(w_{ij}) \neq 0$, and refer to these networks as sparse unbalanced random networks obeying Dale's law.
        
\begin{figure}
                \centering
                %\subfloat[]{%
                    \includegraphics[clip,width=0.87\columnwidth]{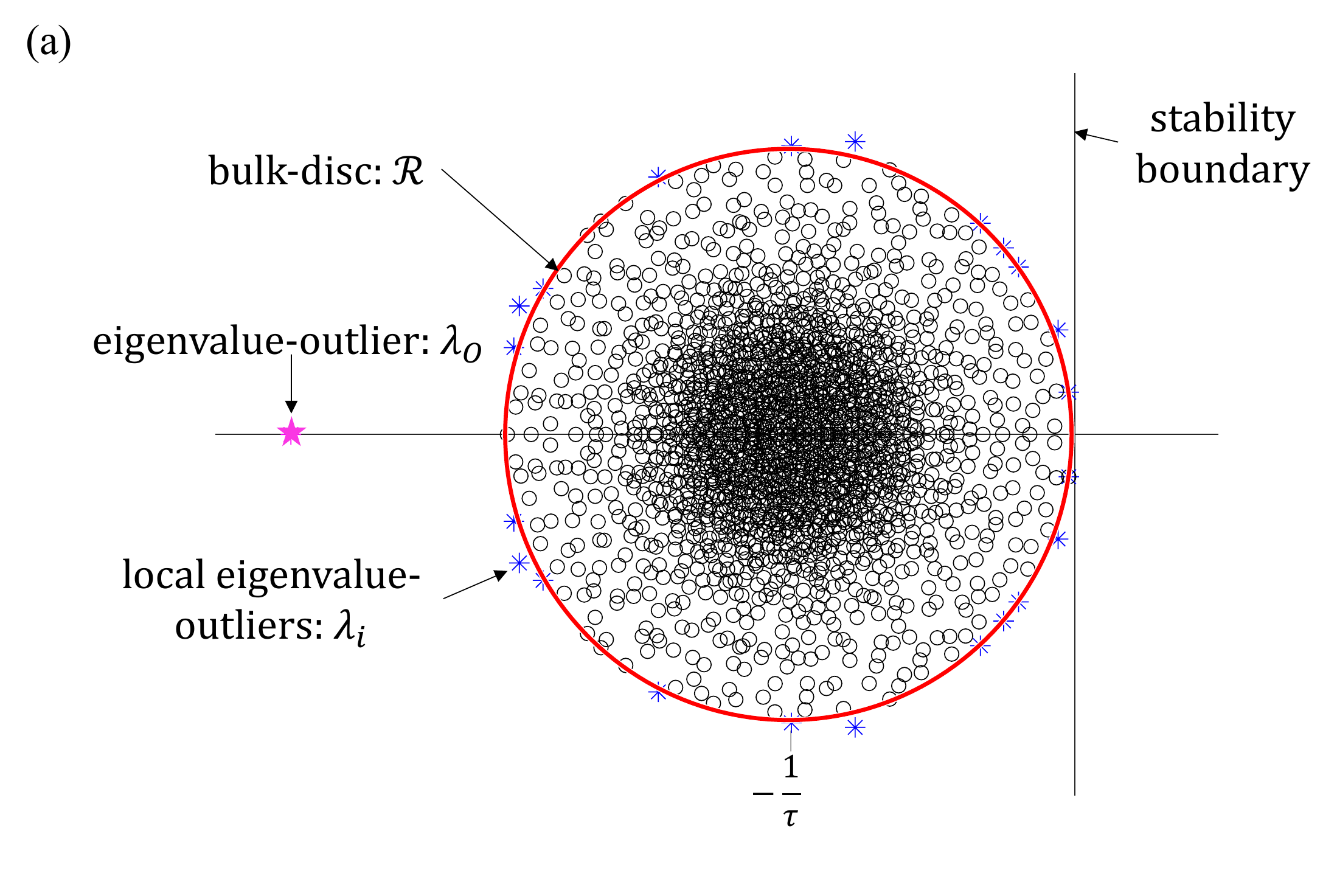}%
                %}
                \quad
                \label{fig:twopopfigORa}
                
                %\subfloat[]{%
                    \includegraphics[clip,width=0.92\columnwidth]{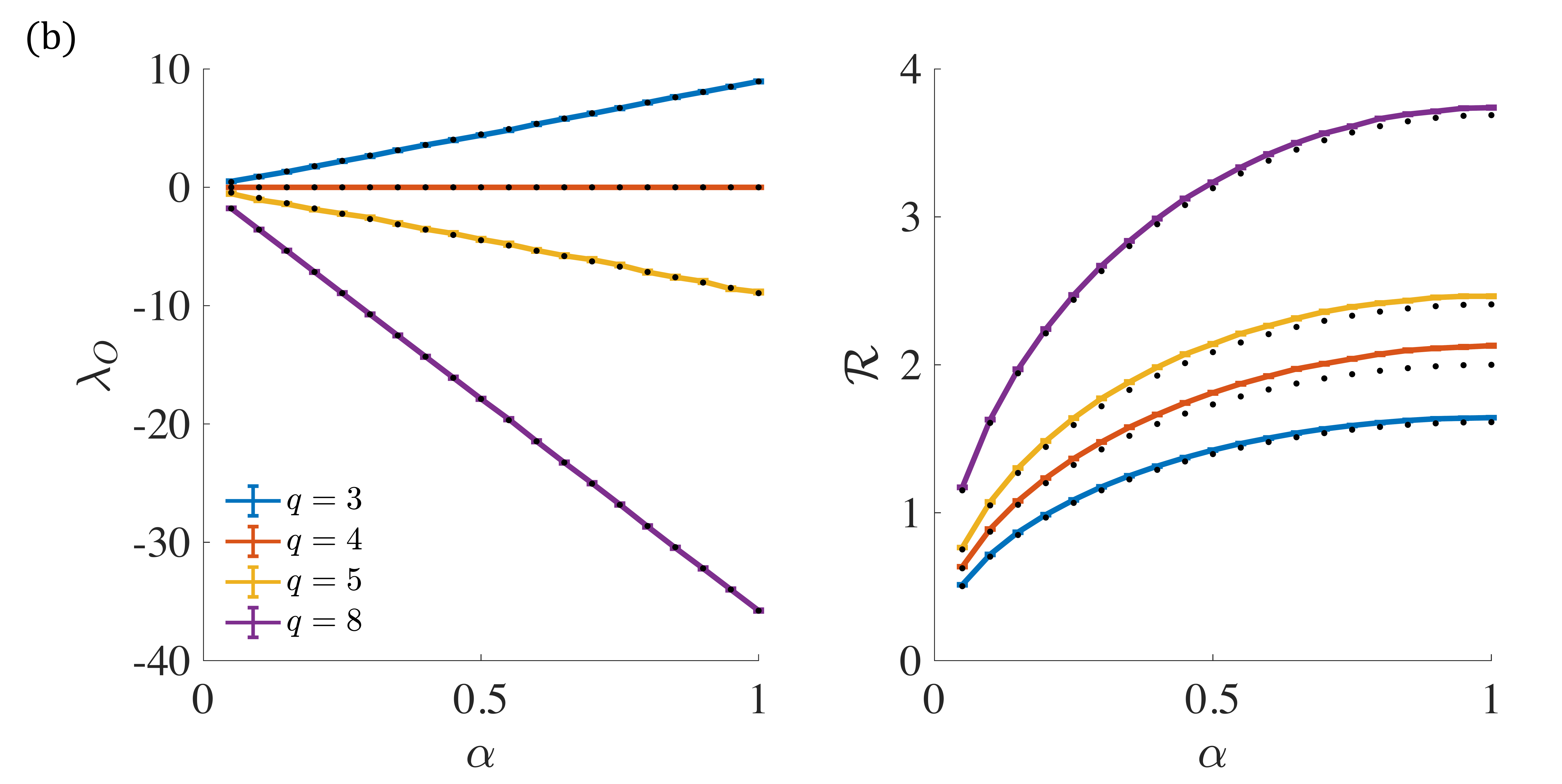}%
                %}
                \quad
                %\subfloat[]{%
                    \includegraphics[clip,width=0.92\columnwidth]{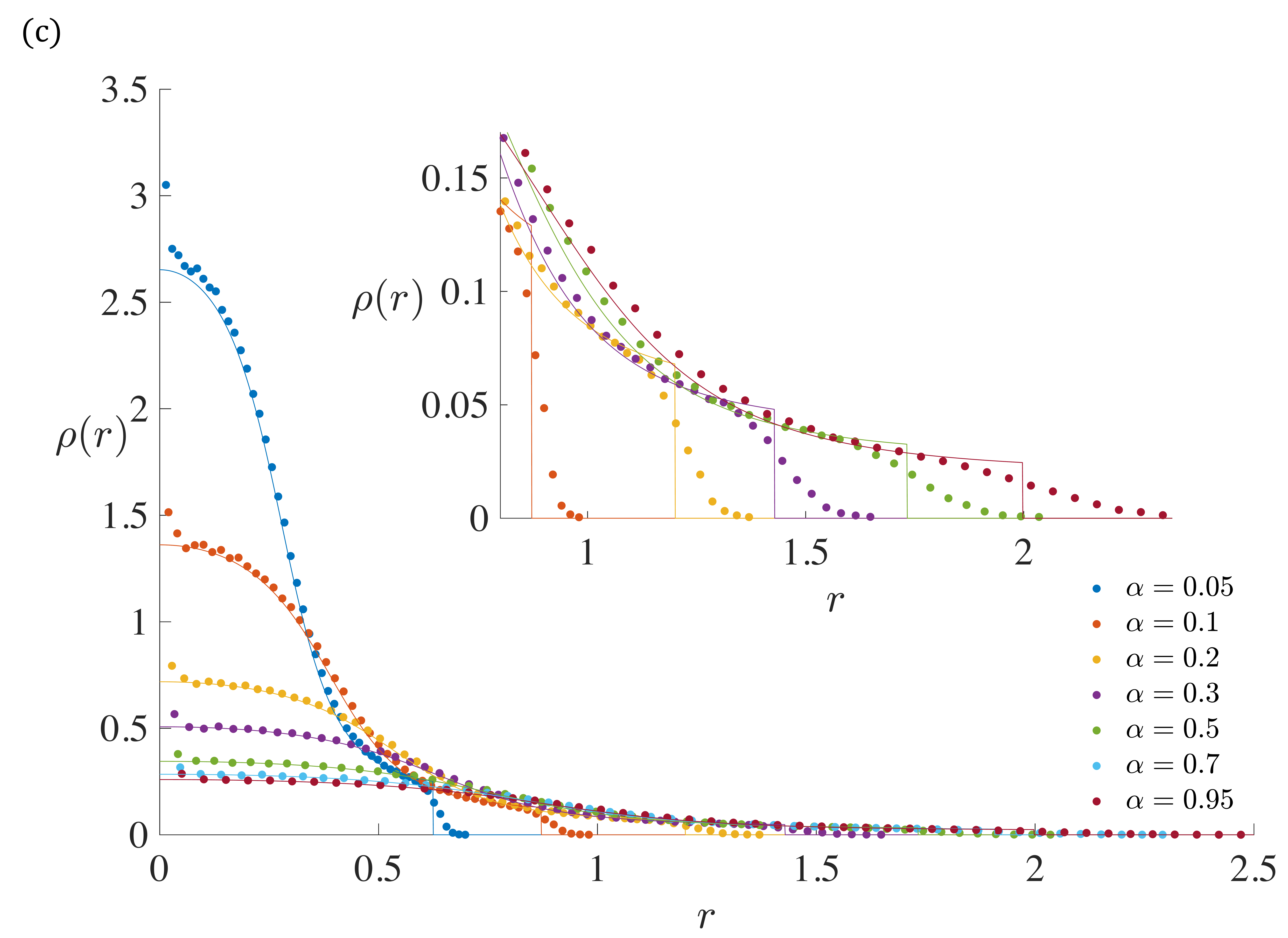}%
                %}
                \quad
            \caption{(a) Eigenvalue spectrum of $W$ with: $\mu_{\mathrm{i}} = -\frac{4.7}{\sqrt{N}}, \sigma_{\mathrm{i}} = \frac{4.7}{\sqrt{N}}, \alpha = 0.99$. (b) Eigenvalue outlier (left) and radius of the eigenspectral-disc (right) of $W$, plotted against $\alpha$ for $\mu_{\mathrm{i}} = -\frac{q}{\sqrt{N}}$, $\sigma_{\mathrm{i}} = \frac{q}{\sqrt{N}}$. 
            The theoretical outliers and radii were calculated using Eqs.~\ref{equ:lambdao2}, \ref{equ:rad2a}, respectively for different ratios, $q$, of inhibition to excitation (blue, orange, yellow, purple). The outlier $\lambda_o$ is zero for the balanced case (orange) and was theoretically calculated (black) using Eq.~\ref{equ:lambdao2} and numerically (red) from $W$. $\mathcal{R}$ was calculated theoretically (black) using Eq.~\ref{equ:rad1b} and numerically (red) by the average second largest eigenvalue of $W$. Numerical computations were averaged over 100 realisations of $W$ and are shown with standard error bars. (c) Spectral density plotted against disc radius for different sparsity, $\alpha$, for theoretical solid lines and numerical points averaged over 500 realisations of $W$ for $\mu_{\mathrm{i}} = -\frac{4}{\sqrt{N}}, \sigma_{\mathrm{i}} = \frac{4}{\sqrt{N}}$.  The inset figure zooms in $\rho(r)=[0, 0.15]$ as the eigenvalue spectral radius is crossed for different levels of sparsity. Note the difference between the analytical (lines) and numerical (dots) density drop offs, indicating that a small number of eigenvalues escape the spectral disc radius. . All plots used parameters of $W$: $N = 2000, f=0.8$, $\mu_{\mathrm{e}} = \frac{1}{\sqrt{N}}$, $\sigma_{\mathrm{e}} = \frac{1}{\sqrt{N}}$.} \label{fig:twopopfigOR}
            \end{figure}

        A typical eigenspectrum for a sparse unbalanced random matrix $W$ that obeys Dale's law is shown in Figure~\ref{fig:twopopfigOR}(a). This figure shows the eigenvalue outlier, $\lambda_O$, and the radius of the eigenspectral-disc, $\mathcal{R}$, as the two primary properties of the eigenspectral distribution. These properties are hypothesised to be defined by $\lambda_O = \EX(w_{ij}) N$, and $\mathcal{R} = \sqrt{\VAR(w_{ij})N}$ \cite{rajan2006eigenvalue,tao2008random, tao2013outliers}.  Note that in this case not all eigenvalues will lie within the radius of the eigenspectral-disc, a few local outliers are located outside the disc radius . In the fully connected case these outliers are controlled by a Zero Row-Sum (ZRS) condition \cite{rajan2006eigenvalue,tao2013outliers} which we extend to the sparse case and discuss later. 
        
        We commence as before by deriving expressions for the mean and variance of the weights $w_{ij}$ in $W$, where $W$ is defined by Eq.~\ref{equ:sWrankperturb}. The mean and variance are given by,
        \begin{align}
            \EX(W)&= f  \mu_{\mathrm{se}} + (1-f) \mu_{\mathrm{si}} \label{equ:ex2mu},\\
            \VAR(W) &= f\sigma_{\mathrm{se}}^{2} + (1-f) \sigma_{\mathrm{si}}^{2} \label{equ:ex2sig},
        \end{align}
        where
        \begin{align}
            \mu_{\mathrm{sk}} &= \alpha \tilde{\mu}_{k} \label{equ:snpm}\\
            \sigma_{\mathrm{sk}}^{2} &= \alpha (1-\alpha) \tilde{\mu}_{k}^{2} + \alpha \tilde{\sigma}_{k}^{2} \label{equ:snpv}
        \end{align}
        are the means and variances of the excitatory and inhibitory weights ($k=\mathrm{e}, \mathrm{i}$) derived in Appendix~\ref{app:derivation2}.
        
        Therefore, the location of the eigenvalue-outlier, and approximate (not including the local outliers, discussed later in Section~\ref{sec:zrsc}) radius of the central eigenspectral-disc for sparse unbalanced random matrices that obey Dale's law can be expressed as, 
        \begin{align}
            \lambda_{O} &= N\left[f  \mu_{\mathrm{se}} + (1-f) \mu_{\mathrm{si}}\right], \label{equ:lambdao2}\\
            \mathcal{R} &= \sqrt{N \left[f\sigma_{\mathrm{se}}^{2} + (1-f) \sigma_{\mathrm{si}}^{2}\right]} \label{equ:rad2a}.
        \end{align}
        We compare our theoretical eigenvalue outlier values and radii to numerical eigenspectra of sparse random matrices constructed using Eq.~\ref{equ:sWrankperturb}. This comparison of the eigenvalue outlier values and the radius of the central eigenspectrum disc is shown in Figure~\ref{fig:twopopfigOR}(b) (and Figure~\ref{fig:twopopmeansvarOR}(a-b) in Appendix~\ref{app:additionalfigures}). The predicted eigenvalue outlier location and radius of eigenspectral-disc radius have an excellent agreement with the numerically calculated eigenvalue outlier and radius of the eigenspectral-disc. 
        
        We observe that sparsity appears linearly in the expression for the degree of structural E-I balance, and hence the eigenvalue outlier Eq.~\ref{equ:lambdao2}. Additionally, the radius of the eigenspectral disc is non-linearly dependent on the sparsity parameter, $\alpha$, the normalised population means $\tilde{\mu}_{\mathrm{e}}, \tilde{\mu}_{\mathrm{i}}$, and normalised population variances $\tilde{\sigma}_{\mathrm{e}}^{2}, \tilde{\sigma}_{\mathrm{i}}^{2}$. From Eq.~\ref{equ:rad2a} we see that for fully connected matrices $\alpha=1$, the radius depends only on the population variances, a result consistent with \cite{rajan2006eigenvalue,tao2013outliers, ipsen2020consequences}. However, once sparsity is introduced into the connectivity, $0<\alpha < 1$, the eigenspectral disc radius changes as a function of all population statistics ($\alpha, \mu_k, \sigma_k^2$). 
    
        \subsubsection{Non-uniform spectral density of eigenvalue distribution} \label{sec:sdf}
        Previous studies  \cite{rajan2006eigenvalue,tao2013outliers,ipsen2020consequences} have shown that a difference in the variances of the excitatory and inhibitory weights, $\sigma_{\mathrm{e}}^{2}, \sigma_{\mathrm{i}}^{2}$, cause the density of the eigenvalue distribution to be non-uniform. We extend these results to the sparse case and show that it is the difference in variances of the sparsely connected excitatory and inhibitory weights, $\sigma_{\mathrm{se}}^{2}, \sigma_{\mathrm{si}}^{2}$ that causes the density to be non-uniform. This phenomena is observed in Figure~\ref{fig:twopopfigOR}(a~\&~c).
        
        The central result of \cite{rajan2006eigenvalue} is the derivation of the expression of the non-uniform eigenspectral density for connectivity matrices obeying Dale's law. The density is a function of the distance to the centre of the disc on the complex plane, $|z|$, and the expression was simplified further in~\cite{ipsen2020consequences}. We incorporate our expressions for the sparse neural population variances Eq.~\ref{equ:snpv} into the spectral density expression defined in~\cite{ipsen2020consequences}. 
        
        The global spectral density for a sparse (un)balanced random connectivity matrix obeying Dale's law is
            \begin{equation}
                \rho_{RA} (z) = \begin{cases}
                        \frac{1}{\pi N \sigma_{\mathrm{si}}^{2}} \left[1-\frac{g}{2} \mathcal{H}_{f}\left(g \frac{|z|^2}{N \sigma_{\mathrm{si}}^{2}}\right)\right]& | z | \leq \mathcal{R} \\
                        0 & | z | > \mathcal{R}
                    \end{cases} \label{equ:sdf}
            \end{equation}
            with 
            \begin{equation}
                g = 1-\sigma_{\mathrm{si}}^{2}/\sigma_{\mathrm{se}}^{2} = 1- \frac{(1-\alpha)\mu_{\mathrm{i}}^{2} + \sigma_{\mathrm{i}}^{2}}{(1-\alpha)\mu_{\mathrm{e}}^{2} + \sigma_{\mathrm{e}}^{2}}. \label{equ:gsdf}
            \end{equation}
            and
            \begin{equation}
                \mathcal{H}_f(x) = \frac{2f - 1 + x }{\sqrt{1 + x(4f-2+x)}} + 1. \label{equ:Hf}
            \end{equation}
        By symmetry an equivalent expression holds for $\sigma_{\mathrm{si}}^{2}$ and $\sigma_{\mathrm{se}}^{2}$ interchanged in Eq.~\ref{equ:sdf} and Eq.~\ref{equ:gsdf}, with $2f-1$ replaced by $1-2f$ in Eq.~\ref{equ:Hf}. Note that the conditions on $\sigma_{\mathrm{sk}}^{2}$ for $k=\mathrm{e}, \mathrm{i}$, and $g$ stated in \cite{ipsen2020consequences} are not required, as the symmetrical expressions yield the equivalent outputs regardless of whether these conditions are met.
        
        Using this expression, the spectral density curves for a sparse balanced and unbalanced random connectivity matrices obeying Dale's law are calculated and compared with numerically simulated density curves for the same parameters. The agreement between the analytical expression and the numerical simulation is shown in Figure~\ref{fig:twopopfigOR}(c) (and Figure~\ref{fig:twopopmeansvarOR}(c) in Appendix~\ref{app:additionalfigures}). 
        As expected, the results indicate that the spectral density curves are dependent on the variances of the sparse excitatory and inhibitory weights $\sigma_{\mathrm{si}}^{2}$ and $\sigma_{\mathrm{se}}^{2}$. Consequently, the density curves are also dependent on the sparsity parameter $\alpha$, the population means $\mu_{\mathrm{e}}, \mu_{\mathrm{i}}$, and population variances $\sigma_{\mathrm{e}}^{2}, \sigma_{\mathrm{i}}^{2}$. In particular Eq.~\ref{equ:gsdf} highlights the conditions for which the density will be uniform, i.e., $g=0$. Previously, the density was only non-uniform if the neural population variances were not equal, $\sigma_{\mathrm{e}}^{2} \neq \sigma_{\mathrm{i}}^{2}$ \cite{rajan2006eigenvalue,ipsen2020consequences}. However, by introducing sparsity, a uniform density ($g=0$) becomes the special case, when the population means and variances obey $|\mu_{\mathrm{e}}|=|\mu_{\mathrm{i}}|$, and $\sigma_{\mathrm{e}}^{2} = \sigma_{\mathrm{i}}^{2}$, respectively. Every other case for non-fully connected networks now has a non-uniform density. 
        \\ \\
        %condition for a uniform spectral density ($g=0$) is much more specific. It requires that sparse population variances be equal $\sigma_{ke}^{2} = \sigma_{ki}^{2}$. This implies that the simplest solution ($g=0$) is equal population means 

        We reformulate the expression for the spectral density such that the formula is symmetric with respect to the two variances. Let $P_{\mathrm{sk}} = \frac{1}{\sigma_{\mathrm{sk}}^{2}}$ for $k=\mathrm{e}, \mathrm{i}$ be the precision of the excitatory and inhibitory weight distributions, respectively. Hence we define the spectral density in terms of the sum of the precisions, $\Sigma P = P_{\mathrm{se}} + P_{\mathrm{si}}$, the difference in the precisions, $\Delta P = P_{\mathrm{se}} - P_{\mathrm{si}}$, and the difference of proportions, $\Delta f = 2f-1$, as
        \begin{equation}
                \rho (z) = \begin{cases}
                        \frac{1}{2 \pi N} \left[\Sigma P - \Delta P \mathcal{H}_{\Delta f}\left(\Delta P |z|^2\right)\right]& | z | \leq \mathcal{R} \\
                        0 & | z | > \mathcal{R}
                    \end{cases} \label{equ:sdf2}
        \end{equation}
        where, 
        \begin{equation}
            \mathcal{H}_{\Delta f}(x) = \frac{x - \Delta f N}{\sqrt{(x -\Delta f N)^2+ N^2(1-\Delta f^2)}}.  \label{equ:HDelf}
        \end{equation}
        The differences in precisions $\Delta P$ and proportion $\Delta f$ can be switched around to favour inhibition, and the expression is equivalent. This new formulation emphasises that the non-uniformity is linked to the $|z|^2$ term, which is paired only with the difference in the precisions $\Delta P$. Therefore, it is $\Delta P$, that cause the non-uniform spectral density. By definition $\Delta P$ depends on the sparsity parameter, and the mean and variance of the excitatory and inhibitory weights, and hence so does the spectral density. This reformulation gives a detailed insight into the interaction of the statistics with the density of eigenvalues across the disc.
            
    \subsubsection{Local eigenvalue-outliers: a zero row-sum (ZRS) condition.} \label{sec:zrsc}
\begin{figure}
            \centering
            %\subfloat[]{%
              \includegraphics[clip,width=0.95\columnwidth]{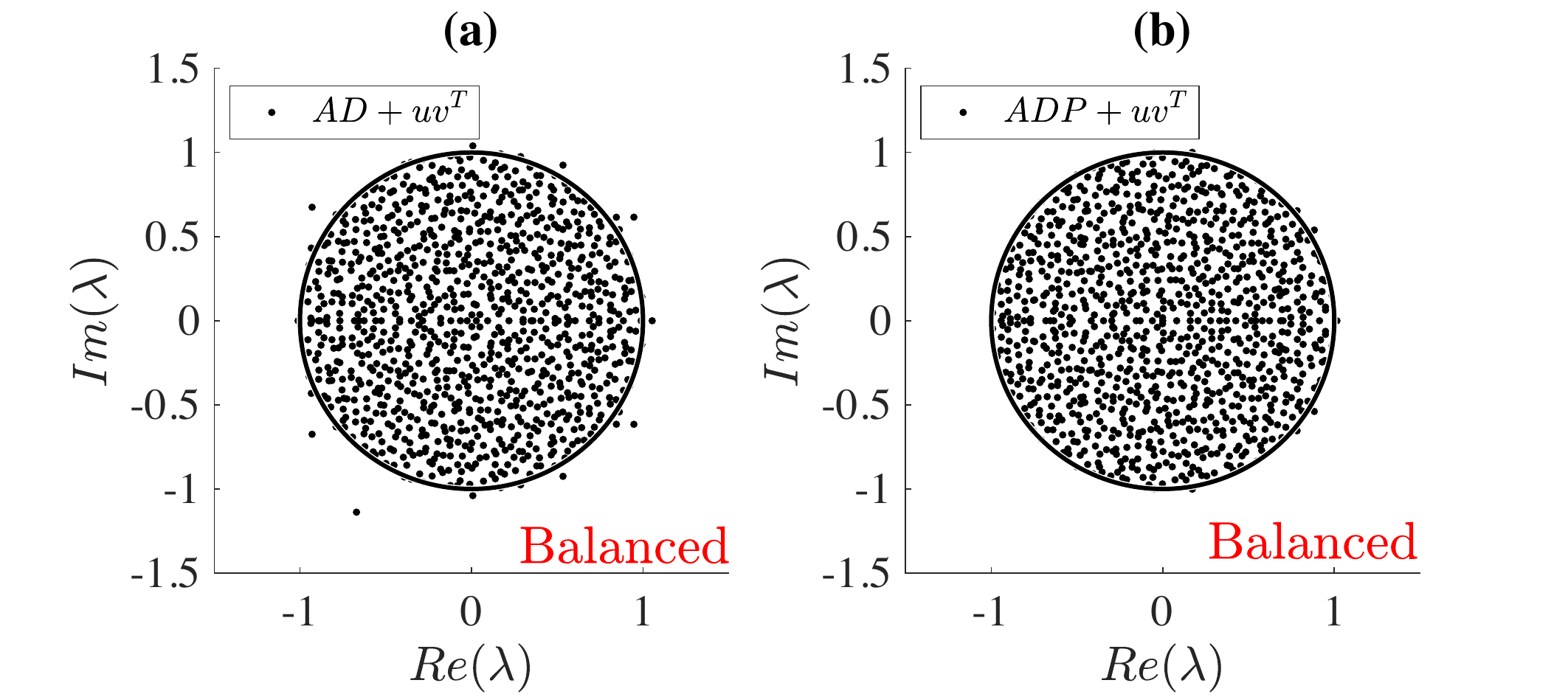}%
            %}
            \quad
            %\subfloat[]{%
              \includegraphics[clip,width=0.95\columnwidth]{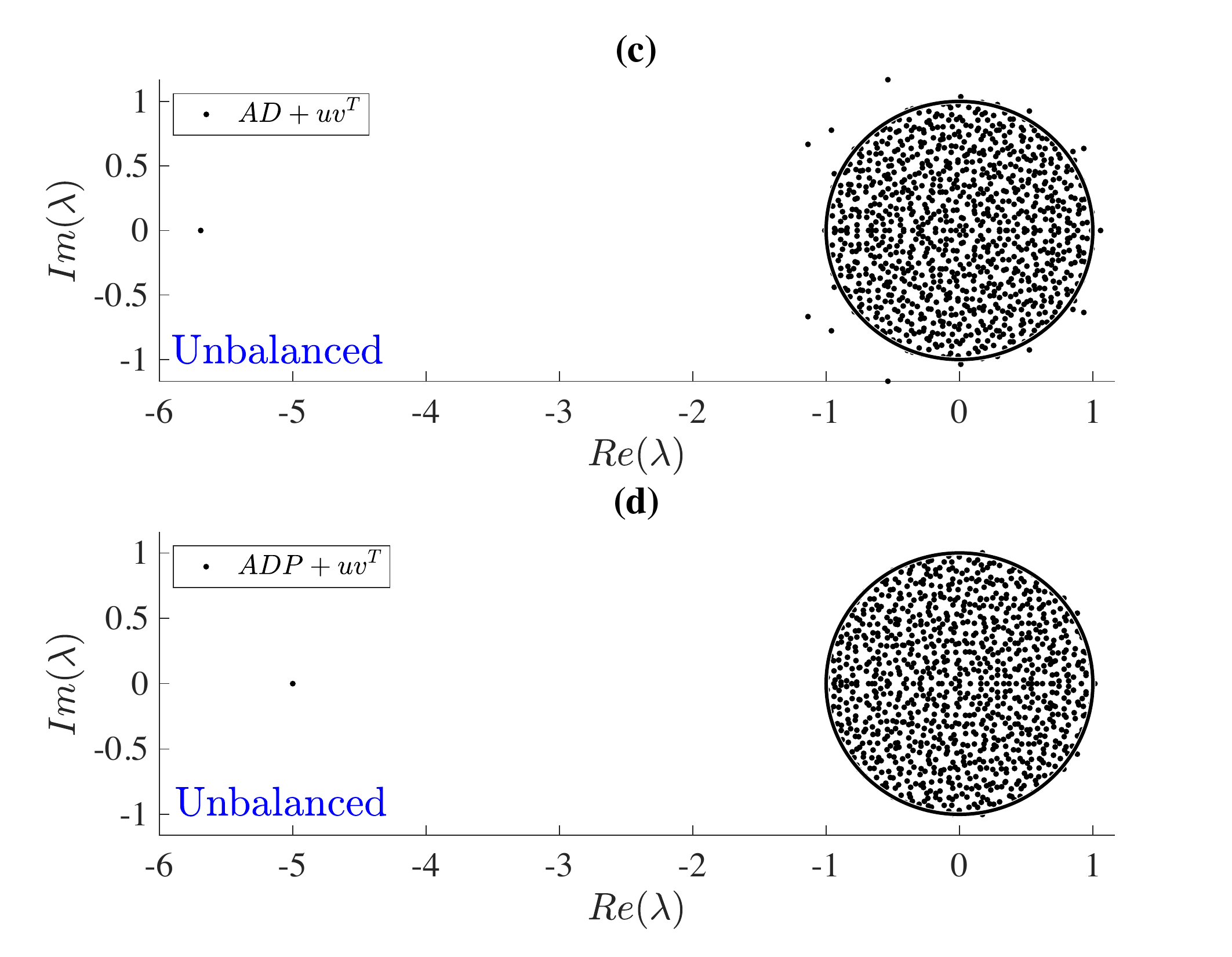}%
            %}
            \caption{Eigenvalue spectra plots of $W$ with $N=100$ and $D = \frac{1}{\sqrt{N}} \mathbb{I}_N$ for (a-b) the balanced case and (c-d) the unbalanced case. (a) Eigenspectra of the matrix $AD+ \boldsymbol{uv}^{\top}$ and (b) with projection operator applied $ADP + \boldsymbol{uv}^{\top}$ to ensure a ZRS. When ZRS is applied the eigenvalues of $ADP$ and $ADP + \boldsymbol{uv}^{\top}$ are identical with no local eigenvalue outliers. (c) Eigenspectra for the unbalanced matrix $AD + \boldsymbol{uv}^{\top}$ and (d) with projection operator applied $ADP + \boldsymbol{uv}^{\top}$. Note that in the balanced case the projection operator annihilates the deterministic term $\boldsymbol{uv}^\top$ only acting on the random term $ADP$ within $W$. However, in the unbalanced case the projection operator $P$ is \textit{only} applied to the random component $ADP$. This distinction is made because if $P$ is applied to the entire connectivity matrix $W$ in the unbalanced case, then $P$ enforces a zero-row sum but also removes the imbalance imposed by $\boldsymbol{uv}^\top$.}
            \label{fig:fullconnectedzrs}
        \end{figure}

        In both the sparse and fully connected cases, we observe a small number of local eigenvalue outliers escaping the circular support Figure~\ref{fig:fullconnectedzrs}(a). These eigenvalue crossings have been previously studied for fully connected \underline{balanced} random matrices obeying Dale's law \cite{rajan2006eigenvalue, tao2013outliers, ipsen2020consequences}. We extend the analysis here firstly to fully connected unbalanced random matrices obeying Dale's law and then to the sparse case.
        
       To control these eigenvalue outliers, previous work \cite{rajan2006eigenvalue, tao2013outliers, ipsen2020consequences} defined a projection operator to ensure that the row-sum of the synaptic connectivity matrix is zero, referred to as the zero row sum (ZRS) condition. This condition ensures that in the thermodynamic limit all eigenvalues converge to lie within the circular support radius. 

        For fully connected balanced random matrices obeying Dale's law, the ZRS condition implemented through a projection operator $P$ is defined as \cite{tao2013outliers,rajan2006eigenvalue,ipsen2020consequences}
        \begin{equation}
            P = \mathbb{I}_N - \frac{\boldsymbol{uu}^\top}{N}, \label{equ:Oprojection}
        \end{equation}
        with $u=(1,\ldots,1)^\top$. The operator $P$ is used such that the synaptic connectivity matrix takes the form, 
        \begin{equation}
            W = A D P + \boldsymbol{uv}^\top, \label{eq:Ozrs}
        \end{equation}
         with $\boldsymbol{v}$ defined by Eq.~\ref{equ:uv} . In \cite{tao2013outliers,rajan2006eigenvalue} the operator $P$ is applied to the entire connectivity matrix $W$, but by construction $\boldsymbol{v}^\top P =\boldsymbol{v}^\top$ in the balanced case, as $\boldsymbol{v}^\top \boldsymbol{u}= 0$, so $P$ need only be applied to the first term. It has been shown in this case that the bounded rank perturbation $\boldsymbol{uv}^\top$ has no effect on the eigenvalues within the circular disc, i.e., the matrix $ADP + \boldsymbol{uv}^\top$ and $ADP$ have identical eigenvalues, \cite{tao2013outliers,rajan2006eigenvalue}. 

        In the unbalanced case, if the projection operator is applied to the entire connectivity matrix $W$ then, by construction, $P$ enforces a zero-row sum but also removes the imbalance imposed by $\boldsymbol{uv}^\top$. However, imbalance can be retained if $P$ is applied to \textit{only} the random component $A D$, i.e., $W = ADP + \boldsymbol{uv}^\top$. Then the argument in \cite{rajan2006eigenvalue} may be extended as follows to show that all eigenvalues of $ADP + \boldsymbol{uv}^\top$ are the same as those $ADP$, except for the outlier eigenvalue from the rank-1 perturbation. Further these shared eigenvalues converge to lie within the circular disc \cite{tao2013outliers} in the thermodynamic limit.
        
        Imposing the projection operator on the random component, $ADP$, of the matrix $W$ ensures that
        \begin{equation}
            ADP \boldsymbol{u} = \boldsymbol{0},  \label{equ:rightevec}
        \end{equation}
        as $ P \boldsymbol{u} = \boldsymbol{0}$. So $\boldsymbol{u} $ is a right eigenvector of $ADP$ with eigenvalue $\lambda = 0$. But $\boldsymbol{u}$ is also a right eigenvector of the full matrix $W$ as 
        \begin{equation}
            (ADP +\boldsymbol{uv}^\top) \boldsymbol{u} =  \boldsymbol{0} + \boldsymbol{uv}^\top \boldsymbol{u} = (\boldsymbol{v \cdot u}) \boldsymbol{u}. \label{equ:rightevec}
        \end{equation}
        Essentially, by adding $\boldsymbol{uv}^\top$ to $ADP$, the eigenvalue $\lambda = 0$ of $ADP$ is changed to $\lambda_{O} = \boldsymbol{v \cdot u} =N(f\tilde{\mu}_{\mathrm{e}}+(1-f)\tilde{\mu}_{\mathrm{e}})$, with corresponding right eigenvector $\boldsymbol{u}$.  
        
       The remaining eigenvalues $\lambda_k$ of $ADP$ are not equal to 0, almost surely (i.e. with probability 1)  \cite{tao2013outliers}. Therefore, for the corresponding left eigenvector, $\boldsymbol{l}_k$, we may substitute the rearranged eigenvalue equation, $\boldsymbol{l}_k =  \frac{\boldsymbol{l}_k ADP}{\lambda_k}$, to find 
        \begin{equation}
            \boldsymbol{l}_k \boldsymbol{uv}^\top =  \frac{\boldsymbol{l}_k ADP}{\lambda_k}\boldsymbol{uv}^\top = \boldsymbol{0}, \label{equ:evalJ}
        \end{equation}
        using Eq.~\ref{equ:rightevec}. Therefore, we have 
        \begin{equation}
            \boldsymbol{l}_k (ADP + \boldsymbol{uv}^\top)  =  \lambda_k\boldsymbol{l}_k .  \label{equ:evalM}
        \end{equation}
        As a result, both matrices $ADP$ and $ADP+\boldsymbol{uv}^\top$ have identical eigenvalues $\lambda_k \neq 0$. Consequently, we conclude in the unbalanced case that the projection operator ensures a zero row sum condition (ZRS) on $ADP$ and therefore controls all eigenvalue outliers, except for the single eigenvalue located at $\lambda_{O} = \boldsymbol{v \cdot u}$. This is what is observed empirically in Figure~\ref{fig:fullconnectedzrs}(c,d).

        %By applying the projection operator to obtain the matrix $ADP$, a null space for the matrix is introduced and a single eigenvalue at zero appears, i.e., $(ADP)\boldsymbol{u} = 0\boldsymbol{u}$. Adding the unbalanced deterministic term to the random matrix, $ADP + \boldsymbol{uv}^\top$ moves the eigenvalue at zero. 
        % From the eigenvalue equation we have, 
        % \begin{equation}
        %     (ADP + \boldsymbol{uv}^\top)\boldsymbol{u} = 0\boldsymbol{u} + \boldsymbol{u}(\boldsymbol{v}^\top \boldsymbol{u}) = (f\mu_i + (1-f)\mu_e)\boldsymbol{u}, 
        % \end{equation} 
        % where $\boldsymbol{u} = (1, ..., 1)^\top$ is an eigenvector.
        % Hence, the outlier escapes to the point $[f\tilde{\mu}_i + (1-f)\tilde{\mu}_e]$ on the complex plane. Note that this has not been strictly proven as in \cite{tao2013outliers} for the structurally E-I unbalanced case, however, 
         %We conclude that all other eigenvalues of $ADP + \boldsymbol{uv}^\top$ are identical

        %Specifically, even though the operator $P$ does not ahniliate with ensures that all the eigenvalues lie within the circular disc in the thermodynamic limit. 

        %[Sentence: the ZRS condition works for the unbalanced case to ensure the eigenvalues lie within the circular disc - appears to work for the unbalanced case though this has not been proven. ]

    \subsubsection{Local eigenvalue-outliers: a new Sparse Zero Row-Sum (SZRS) condition for sparse random matrices.}       
        We also observe the phenomena of local eigenvalue-outliers in the sparse case, see Figure~\ref{fig:twopopfigOR}(a). Figure~\ref{fig:twopopfigOR}(b) shows that there exists a small discrepancy between the numerically estimated radius (black dots) and the theoretical eigenspectral-disc radius (coloured lines). This small discrepancy is due to a small number of eigenvalue outliers as seen in Figure~\ref{fig:twopopfigOR}(a). Further, Figure~\ref{fig:twopopfigOR}(c) shows the density of eigenvalues does not drop off as precisely in the numerical estimate as it does in the analytical calculation of the density Eq.~\ref{equ:sdf} and Eq.~\ref{equ:gsdf}. This highlights that the number of these eigenvalue-outliers is relatively small, and that their distance from the circular support increases as $\alpha$ approaches 1. 
        
        To remove these outliers, we implement an analogous ZRS condition to that in the previous section i.e., a Sparse Zero Row-Sum condition (SZRS). We hypothesise that the constraint will ensure that all eigenvalues lie within the theoretical radius in the thermodynamic limit. A potential solution would be to derive an analogous projection operator to ensure the rows of $W$ sum to zero. However, due to the nature of matrix multiplication, such an operator will not preserve the sparsity pattern. To ensure the sparsity pattern is preserved we instead enforce a zero row-sum numerically by subtracting the average of the rows from each nonzero entry in the connectivity matrix $W$. 
        %The constraint subtracts the mean of each of the rows $\bar{J_{i}}$ from the nonzero entries in that row $J_{i}$. 
        This is succinctly defined as,
        \begin{equation}
            W = S \circ (AD + \boldsymbol{uv}^\top) - B \label{equ:zrs1}
        \end{equation}
        where $B_{ij} = S_{ij} \bar{W_{i}}$ and the average of the row
        \begin{equation}
            \bar{W_{i}} = \sum_{j} W_{ij}/\sum_{j} S_{ij}. \label{equ:zrsB}
        \end{equation}
        It is important to note that similar to the projection operator for fully connected balanced matrices, the SZRS condition is applied to both components of $W$. However, due to the introduction of sparsity, both terms $S\circ AD$ and $S \circ \boldsymbol{uv}^\top$ are now random matrices and the SZRS now acts on both components (instead of annihilating with $\boldsymbol{uv}^\top$ like in the fully-connected case). 
        
        The SZRS condition enforces a zero row-sum and strictly preserves the sparsity pattern specific to the realisation of $W$. Further, the implementation is equivalent to the original condition introduced for fully connected balanced matrices \cite{rajan2006eigenvalue, tao2013outliers,ipsen2020consequences}. We observe that by applying the SZRS condition to sparse balanced random matrices obeying Dale's law, the eigenvalues of $W$ converge to lie within the circular disc. 
        
        Similar to the fully-connected unbalanced matrices obeying Dale's law, this condition cannot be applied in the sparse unbalanced case without completely removing the imbalance imposed by $\boldsymbol{uv}^\top$. However, we can still apply the constraint to the first component of $W$ to ensure a partial zero-row sum, i.e., $B_{ij} = S_{ij} \bar{J}_{i}$,  
        \begin{equation}
            \bar{J}_{i} = \sum_{j} J_{ij}/\sum_{j} S_{ij}. \label{equ:zrspartial}
        \end{equation} 
        with $J = AD$, thus preserving the imposed imbalance, and partially controlling the local-outliers. 
        
        In the next section we investigate this phenomenon further by constructing a numerical homotopy to examine how sparsity and Dale's law affect eigenvalue crossings with respect to the eigenspectral-disc theoretical radius. Specifically, we explore the effects of the SZRS condition and partial SZRS condition for the balanced and unbalanced cases, respectively. 
        
    \subsubsection{Construction of a homotopic mapping to illustrate the effects of the sparse zero-row sum conditions}
        \begin{figure}
            \centering
            %\subfloat[]{%
              \includegraphics[clip,width=0.97\columnwidth]{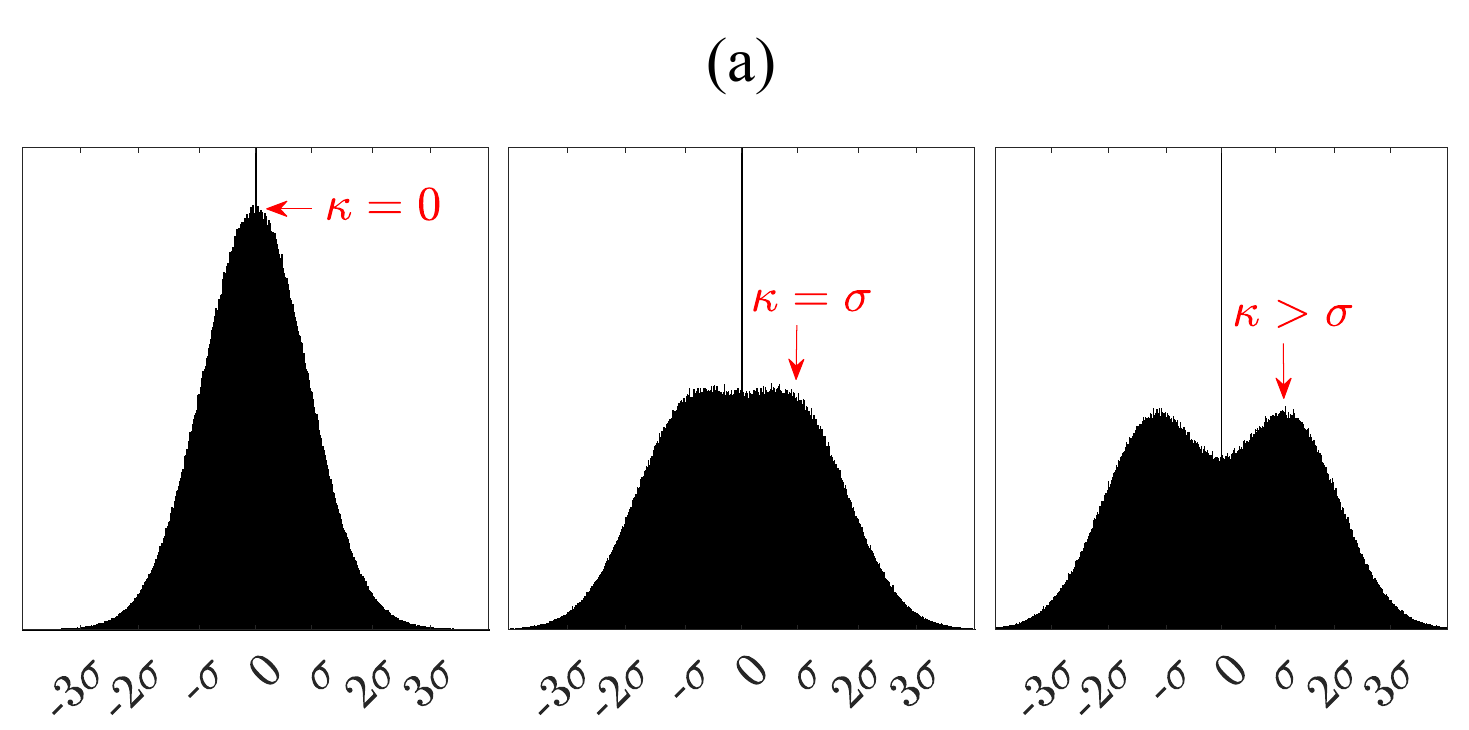}%
            %}
            \quad
            %\subfloat[]{%
              \includegraphics[clip,width=0.95\columnwidth]{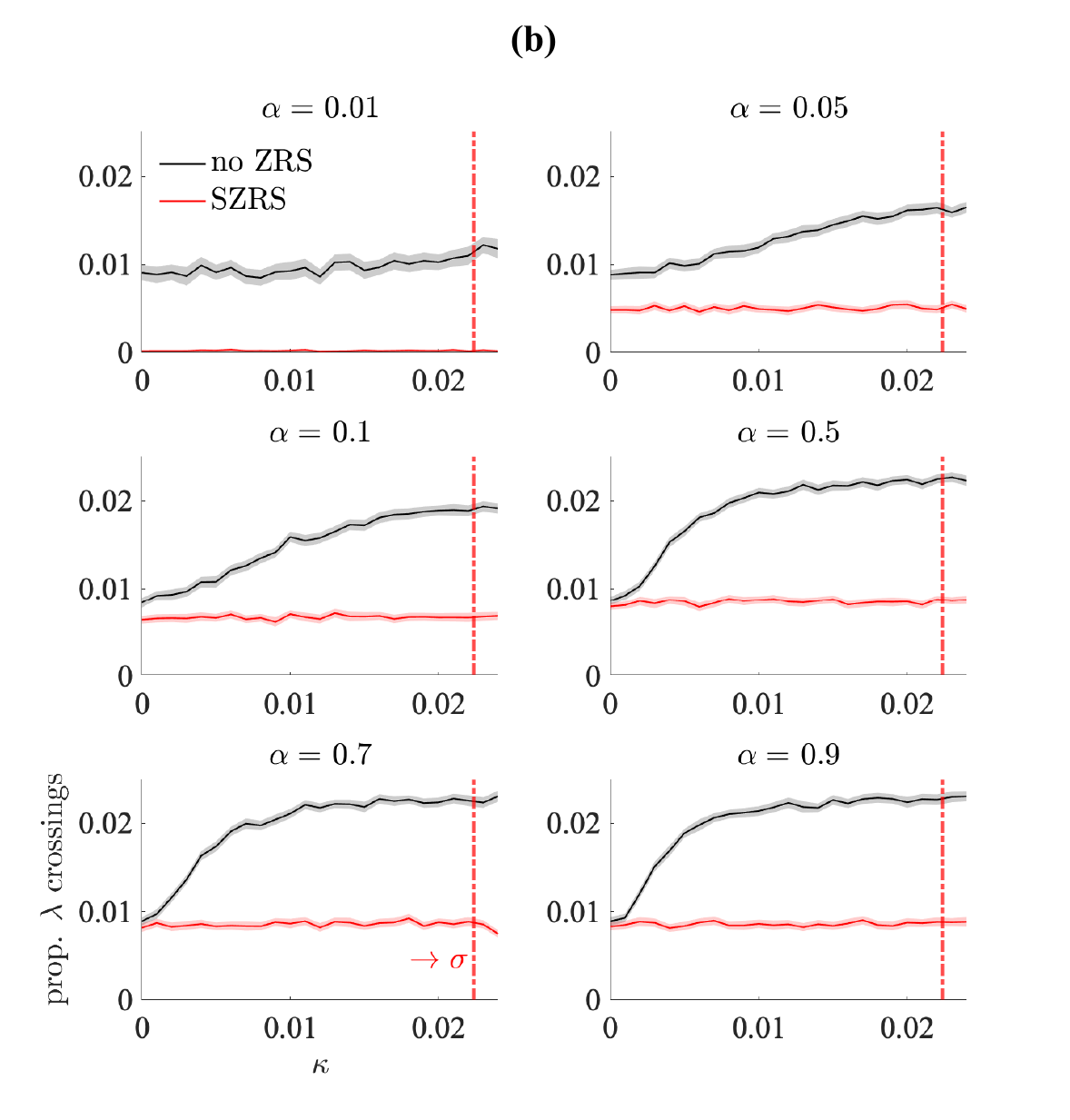}%
            %}
            \quad
            %\subfloat[]{%
              \includegraphics[clip,width=0.95\columnwidth]{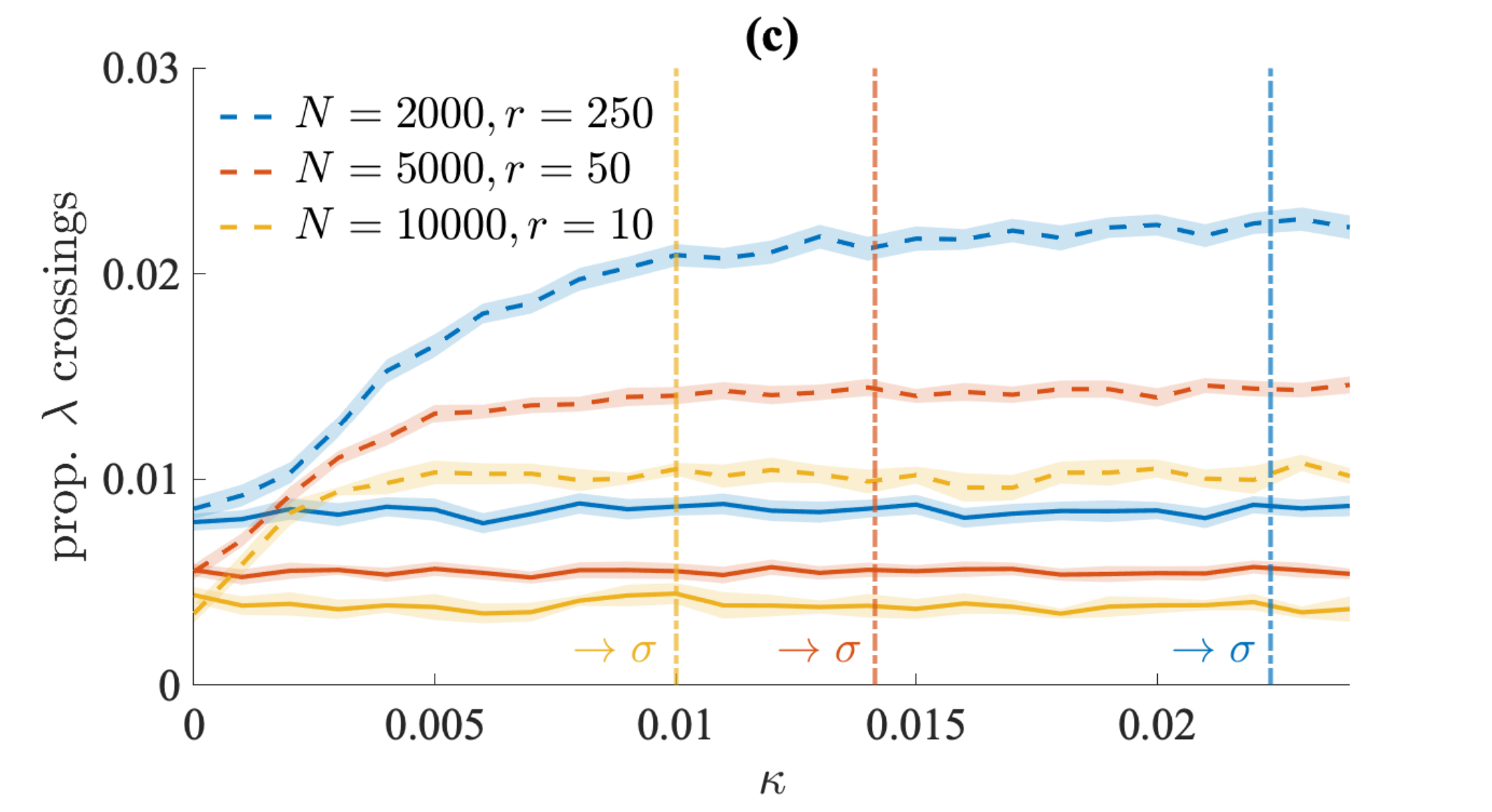}%
            %}
            \caption{(a) Histogram of the distribution of entries in the connectivity matrix $W$. From left to right a single Gaussian distribution ($\kappa=0$) is continuously deformed via $\kappa$ into two distinct Gaussian distributions ($\kappa = \sigma$), with a zero peak due to sparsity.
            (b) Numerical homotopy for the sparse balanced case. Plots show the average proportion of eigenvalue crossings (outliers) as a function of $\kappa=0 \rightarrow \sigma$.  The proportion is an average calculated over $100$ realisations, and the standard error is included as bars on each point.  Plots were calculated with the means of the two populations, $\mu_{\mathrm{e}} = \kappa$, $\mu_{\mathrm{i}} = -\kappa$, and identical population variances $\sigma_{\mathrm{e}}^{2} = \sigma_{\mathrm{i}}^{2} = 1/N = \sigma^2$ for $N=2000$ and $f=0.5$. Each sub-figure shows the homotopy for different values of $\alpha$. (c) Numerical homotopy for the sparse ($\alpha = 0.5$) balanced case indicating finite size effects on the outliers. Solid lines are the average and the shading indicates the standard error. The homotopy is calculated for no ZRS and when SZRS is applied, with above parameter sets with increasing system size $N=2000, 5000, 10000$ and realisations $r=250, 50, 10$, respectively. }
            \label{fig:N2000homotopy}
        \end{figure}
        
        In fully connected random balanced connectivity matrices obeying Dale's law, it is the separation of the means of the two neural distributions that results in eigenvalues crossing the eigenspectral-disc radius to become outliers \cite{rajan2006eigenvalue,tao2013outliers}. We investigate this phenomenon further for sparse random matrices obeying Dale's law. 

        A homotopy is a continuous (but not necessarily homeomorphic) mapping from one limiting case to another limiting case. For example, $H=\kappa F + (1-\kappa)G$ such that as $\kappa$ goes from $0 \rightarrow 1$, where $\kappa$ is the homotopy parameter, then $H$ goes from the function $G$ to the function $F$.  \cite{rahimian2011new}. Specifically, we map how the excitatory and inhibitory neural distributions deform from a single Gaussian distribution ($\mu = \kappa = 0, \sigma = 1/\sqrt(N))$ into two distinct Gaussian distributions with means $\mu_{\mathrm{e}} = \kappa$, and $\mu_{\mathrm{i}} = -\kappa$, and identical variances  $\sigma_{\mathrm{e}}^{2} = \sigma_{\mathrm{i}}^{2} = 1/N = \sigma^2$ . This linear homotopic mapping will show that as the excitatory and inhibitory distributions separate, the proportion of eigenvalues that escape and cross the theoretical eigenspectral disc radius increases.
        
        %To investigate the effects of both sparsity and Dale's law on the eigenvalue-outliers, we perform a numerical homotopic mapping. 
        In this section, we perform the homotopic mapping for two cases i) sparse balanced, and ii) sparse unbalanced random matrices obeying Dale's law. For each case we compare the proportion of crossings when no row-sum condition is implemented and when the SZRS condition or partial SZRS condition is enforced on i) and ii), respectively. 

        The homotopy parameter $\kappa$, defines the degree of separation of the means of the two neural distributions. When $\kappa=0$ the entries $w_{ij}$ of the connectivity matrix form a single Gaussian distribution with a large peak at zero for the sparse case, Figure~\ref{fig:N2000homotopy}(a). The peak at zero changes only with the sparsity parameter $\alpha$. As $\kappa$ increases the Gaussian distribution widens, continuously deforming one population into two populations. As $\kappa \rightarrow \sigma$ the single Gaussian distribution separates into two distinct Gaussian peaks, with a large peak at zero in the sparse case. At this point the means of the excitatory/inhibitory populations are significantly different, i.e., two standard deviations of separation, thus Dale's law is effectively implemented into the connectivity matrix, Figure~\ref{fig:N2000homotopy}(a).

        % explaining figure b
        The numerical homotopy directly demonstrates how this degree of separation, $\kappa$, affects the proportion of eigenvalue crossings for sparse balanced matrices obeying Dale's law. Specifically, Figure~\ref{fig:N2000homotopy}(b) plots the proportion of eigenvalue crossings (averaged over 100 realisations) with and without a SZRS condition i.e. Eq.~\ref{equ:zrs1}, implemented (red), and not implemented (black). In the balanced case, we observe that for all degrees of distribution separation $\kappa$, the SZRS condition (red) ensures that only a very small proportion of eigenvalues cross the disc. In the sparse limit $\alpha \rightarrow 0$ we observe that the number of outliers for matrices with no SZRS condition decreases to a constant value (0.01) for all degrees of separation $\kappa$. Further, with the SZRS condition, the proportion of crossings goes to zero as $\alpha \rightarrow 0$, due to the matrix being closer to an IID distribution. 
        
        Figure~\ref{fig:N2000homotopy}(c) shows that as $N$ becomes larger, the proportion of eigenvalue crossings decreases when the SZRS condition is applied, indicating that they are most likely a finite size effect, consistent with previous results\cite{rajan2006eigenvalue,tao2013outliers}. 

\begin{figure}
            \centering
            %\subfloat[]{%
              \includegraphics[clip,width=0.97\columnwidth]{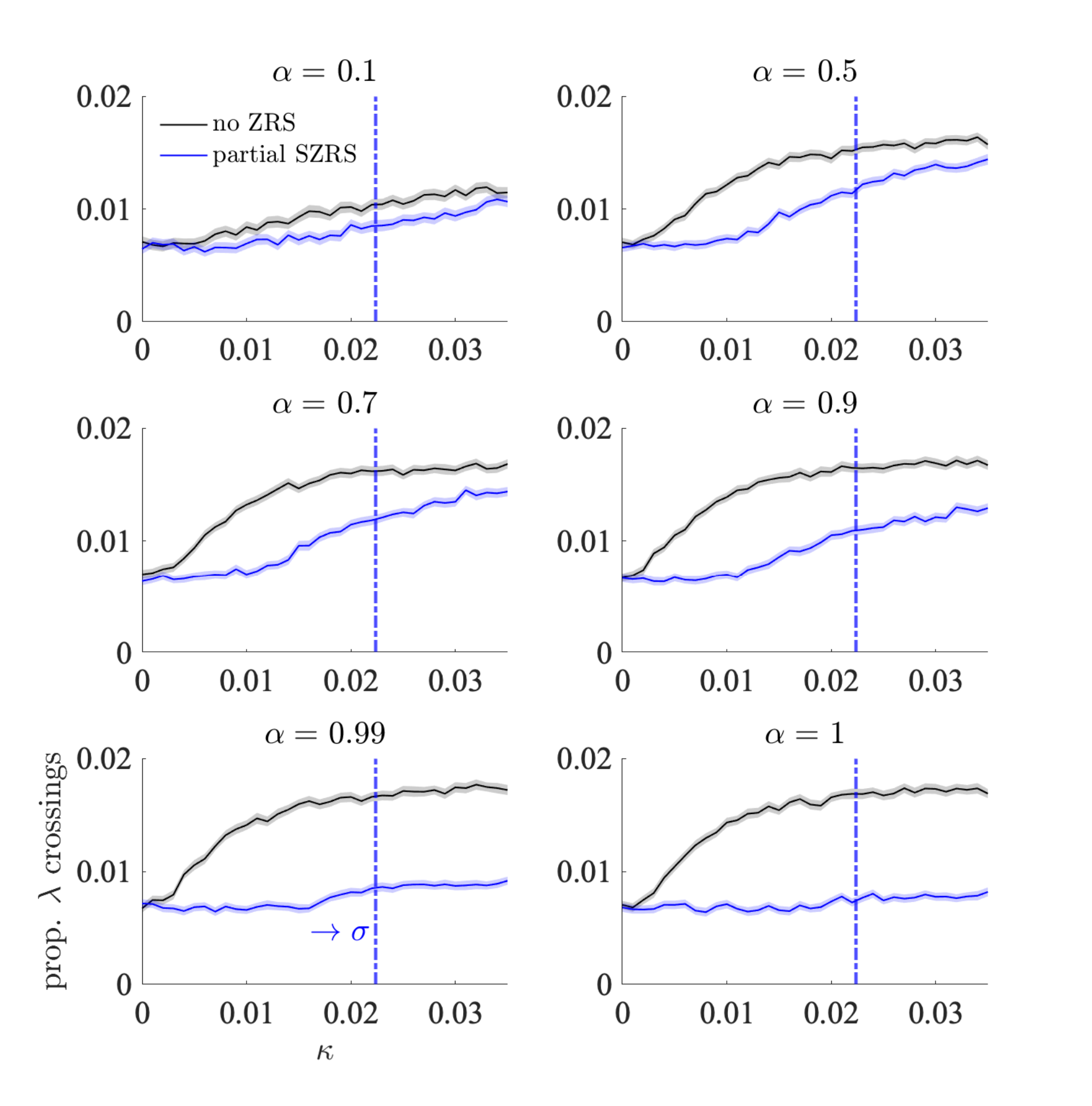}%
            %}
            \caption{Numerical homotopy for the sparse unbalanced case with no ZRS and a partial SZRS applied. Plots shows the average proportion of eigenvalue crossings (outliers) as a function of the continuous deformation from $\kappa=0$ to $\kappa \rightarrow \sigma$. The proportion is an average calculated over $250$ realisations, and the standard error is indicated by the shaded area. The vertical dashed lines denote the point where  $\kappa = \sigma$ and there is sufficient separation between the two distributions such that the excitatory and inhibitory population statistics are distinct. Plots were calculated with $\mu_{\mathrm{e}} = \kappa$, $\mu_{\mathrm{i}} = -5\kappa$, $\sigma_{\mathrm{e}}^{2} = 1/N$, $\sigma_{\mathrm{i}}^{2} = 4^2/N$,  $N=1000$ and $f=0.8$. Each sub-figure shows the homotopy for different values of sparsity $\alpha$. }
            \label{fig:N1000homotopyunbalanced}
        \end{figure}

        Conversely, for the sparse unbalanced case, these outlier eigenvalues cannot be controlled with the SZRS condition without removing the imposed imbalance. Instead, we apply a partial SZRS condition to the unbalanced case. We observe in Figure~\ref{fig:N1000homotopyunbalanced} that for low degrees of distribution separation $\kappa < 0.01$, the partial SZRS condition works most optimally to ensure that only a few eigenvalues cross the eigenspectral-disc radius. Moreover, we find that the partial SZRS condition works better for $\kappa < 0.01$ if the matrix is closer to being fully connected, $\alpha > 0.9$. When $\kappa = \sigma$, the numerical homotopy shows the breakdown of the ability of the partial SZRS condition to minimise the number of eigenvalue outliers. Even when the network is almost fully connected $\alpha = 0.99$,  (see Figure~\ref{fig:N2000homotopy}(b)), the SZRS condition ensures only that there are fewer eigenvalue outliers, and becomes less effective as the network becomes more sparse. 
        
        %escape the circular support. The numerical homotopy analysis illustrates that though the partial SZRS condition works to prevent fewer local eigenvalue-outliers, the condition does not ensure the removal of all local eigenvalue-outliers in the sparse case.

\section{Discussion}
In this paper we examined the stability of random neural networks with more realistic anatomical structures in the form of sparse connectivity and Dale's law. Specifically, we examined the eigenspectrum of the associated network Jacobian for sparse (un)balanced random synaptic connectivity matrices obeying Dale's law. The results presented here consider all levels of network sparseness $0\leq \alpha\leq1$, and also implements Dale's law using distributed weights. Therefore, the expressions derived significantly extend previous studies which only separately considered (i) fully connected random matrices obeying Dale's law \cite{rajan2006eigenvalue,ipsen2020consequences}, (ii) one population sparse random matrices with zero mean \cite{tao2008random,herbert2022impact}, or (iii) sparse matrices (in the sparse limit $\alpha << 1$) with constant weights describing each of the excitatory and inhibitory populations \cite{ostojic2014two,mastrogiuseppe2017intrinsically,shao2023relating}.
    
    \subsection{The distribution of eigenvalues of sparse random matrices.} \par
    The eigenspectrum of the network Jacobian evaluated at the homogeneous equilibrium, consists of an eigenvalue outlier $\lambda_O$ (for the unbalanced case), an eigenspectral-disc with radius $\mathcal{R}$, and non-uniform density of eigenvalues across this disc. We demonstrate that the location of the eigenvalue outlier is linearly related to the sparsity parameter ($\alpha$), and structural (E-I) (im)balance. Interestingly, we show that the variance of the connectivity matrix and consequently the radius of the eigenspectral-disc scales non-linearly with the sparsity parameter and the means and variances of both populations  (Eq.~\ref{equ:rad2a} and Figure~\ref{fig:onepopfigOR}(b)). 
    
    Our results also demonstrate that introducing sparsity to a network of distinct excitatory and inhibitory neural populations changes the spectral density to become non-uniform, even when the population variances are the same. Specifically, by reformulating the spectral density formula in terms of the difference between the precisions Eq.~\ref{equ:sdf2}, we demonstrate that non-uniform density depends not only on the difference in the population variances $\sigma_{\mathrm{e}}^{2} \neq \sigma_{\mathrm{i}}^{2}$, but also the difference in the magnitude of the population means $|\mu_{\mathrm{e}}| \neq |\mu_{\mathrm{i}}|$. This additional requirement is counter-intuitive due to the nonlinear interaction between sparsity $\alpha$ and the population means when $\alpha \neq 1$. Furthermore, if we extend this analysis and define population specific sparsity parameters $\alpha_{\mathrm{e}}, \alpha_{\mathrm{i}}$, then the non-uniform spectral density further depends on this difference as well. These findings are important, because they show an intricate interplay of all of the statistics, rather than just between the variances of the excitatory and inhibitory weights as shown previously \cite{rajan2006eigenvalue,ipsen2020consequences}.

    \subsection{Local outliers, a new ZRSC, and homotopy analysis.} \par
     Our results show that a small number of local eigenvalue-outliers escape the eigenvalue spectral disc radius for sparse random matrices obeying Dale's law (Figure~\ref{fig:twopopfigOR}). Previous works \cite{rajan2006eigenvalue,tao2013outliers} have shown that local eigenvalues escape the disc if the connectivity matrix $W$ is fully connected ($\alpha = 1$) and the means of the excitatory and inhibitory distributions are different. A ZRS condition removes the local eigenvalue-outliers by forcing the eigenvalues of the random component to be the same as those of the connectivity matrix $W$, excluding the eigenvalue outlier that is generated by structural E-I imbalance.
    
    \cite{tao2013outliers} proves that a projection operator for the fully connected balanced case ensures a zero-row sum (ZRS) condition Eq.~\ref{equ:Oprojection}, that forces the local eigenvalue-outliers to converge to lie within the spectral disc in the thermodynamic limit. We extend this proof to the fully connected unbalanced case and show that these eigenvalue-outliers can also be controlled if the projection operator is applied to the random component only. 
    
    In this paper, we also derived and implemented an analogous numerical ZRS condition for sparse balanced random matrices obeying Dale's law Eq.~\ref{equ:zrs1}. The new SZRS condition works effectively to ensure convergence of the eigenvalue-outliers to lie within the disc radius. We note that this condition does not work for the sparse unbalanced case, as not only are the local eigenvalues removed, but also the largest outlier imposed by the imbalance is removed. To retain imbalance, we implement a partial SZRS on the term $S\circ AD$. Unlike the equivalent condition applied to the fully connected case, the partial SZRS condition only ensures that fewer eigenvalues escape the eigenspectrum-disc radius in the thermodynamic limit Figure~\ref{fig:N1000homotopyunbalanced}. We hypothesise that this is due to the fact that the second term $S\circ \boldsymbol{uv}^\top$ in the matrix $W$ is also random and not purely deterministic, as in the fully connected case.

    \subsection{Interaction of sparsity with structural E-I balance} \par 
    Our results show that there exists a fundamental interplay between sparsity and the population means and variances that affect key eigenspectral distribution properties. When Dale's law is implemented in sparse random matrices, the structural (E-I) balance linearly scales with the sparsity parameter $\alpha$ Eq.~\ref{equ:ex2mu}. However, introducing sparsity changes both the variances of the excitatory and inhibitory weights $\sigma_{\mathrm{sk}}^{2}$,($k = e, i$), and the overall variance of the matrix $W$, $\VAR(w_{ij})$. This in turn non-linearly scales the radius and spectral density of the eigenvalue spectral disc. The radius is dependent on the non-linear interaction between the sparsity parameter, $\alpha$, the square of the population means $\mu_{\mathrm{e}}^{2},\mu_{\mathrm{i}}^{2}$, and population variances $\sigma_{\mathrm{e}}^{2},\sigma_{\mathrm{i}}^{2}$ Eq.~\ref{equ:rad2a}. Therefore, unless $\alpha=1$ and~/~or $\mu_{\mathrm{e}}=\mu_{\mathrm{i}}=0$, the radius now depends on the population means and not just the variances as in \cite{rajan2006eigenvalue,ipsen2020consequences}. Interestingly, if the matrix is structurally E-I balanced, i.e. $\EX(W)=0$, the radius of the eigenspectral disc still scales as a function of the sparsity parameter and the population means and variances (see Figure~\ref{fig:twopopmeansvarOR}(b) in Appendix). 
    
    The spectral density is dependent on the differences between the sparse population variances Eq.~(\ref{equ:sdf},\ref{equ:gsdf}), which are non-linearly dependent on the sparsity parameter and the means and variances of both the excitatory and inhibitory populations. The eigenvalue spectral density can be non-uniform even if the network is structurally E-I balanced and the population variances $\sigma_{\mathrm{e}}^2,\sigma_{\mathrm{i}}^2$ are equal. This is due to the dependency of the variances Eq.~\ref{equ:ex2sig} of the excitatory and inhibitory weights on the square of the means $\mu_{\mathrm{e}}^{2},\mu_{\mathrm{i}}^{2}$. The analysis presented in this paper is straightforwardly extendable to distinct sparsity parameters for each neural population, i.e., $\alpha_e \neq \alpha_i$. We observe that this distinction has further implications on the structural E-I balance and hence the eigenvalue outlier, the spectral disc radius, and the spectral density. 
    %Sparsity: If we let the sparsity be different for each population i.e.$\alpha_e \neq \alpha_i$, then from Eq.() even if the population means $\mu_{\mathrm{e}}=-\mu_{\mathrm{i}}$ and the population variances $\sigma_{\mathrm{e}}^{2} = \sigma_{\mathrm{i}}^{2}$, then the spectral density would be non-uniform as well as the radii, structural E-I balance condition and consequently the outlier. 

    \subsection{Implications on neural network dynamics} \par
        Our results provide insight into what combination of factors contribute to the stability of large networks of neurons and other complex networked dynamical systems \cite{allesina2015stability}. Regulating neural function, and dynamic E-I balance must take into account network sparsity at all levels of network connectedness. By examining the eigenspectra of sparse random matrices obeying Dale's law we find that though the eigenspectra are similar to those in their fully connected counterparts \cite{rajan2006eigenvalue}, there are a few key differences that influence the stability, and therefore the transition into spontaneous asynchronous activity. 
        
        \subsubsection{Inducing non-trivial dynamics} \par
        As discussed previously, non-trivial spontaneous asynchronous activity occurs when the eigenspectral-disc crosses the stability boundary. This can happen via the interplay between the time-constant, $\tau$, (which positions the disc centre) and the variance of the connectivity matrix (which determines the radius of the eigenspectral disc) \cite{ipsen2020consequences}. However, for sparse unbalanced random matrices obeying Dale's law this interplay is more complex. 
        
        If the sparse network obeying Dale's law is excitatory dominated, i.e., $\alpha(f\mu_{\mathrm{e}} + (1-f)\mu_{\mathrm{i}})>0$, then the eigenvalue outlier crosses the stability boundary. Destabilisation occurs if the real part of the largest eigenvalue, the eigenvalue outlier $\lambda_O$ is greater than zero. If, however the sparse network is balanced, or inhibitory dominated, $\alpha(f\mu_{\mathrm{e}} + (1-f)\mu_{\mathrm{i}})\leq 0$, non-trivial dynamics are first induced by the radius of the eigenspectral-disc crossing the stability boundary. Previous studies found that the transition is induced solely by the interplay between the membrane time constant and the population variances \cite{rajan2006eigenvalue,mastrogiuseppe2017intrinsically,ipsen2020consequences}. We find that for structural (E-I) balanced and inhibitory dominated networks, the relationship is significantly more complex than found previously. Specifically, the transition depends on the interaction between the membrane time constant, the sparsity parameter, the population means, and the population variances.
        
        Furthermore, the non-uniform eigenspectral density ensures that more eigenvalues lie near the centre of the disc and fewer eigenvalues lie near the edge. This means that there are fewer eigenvalues that lie adjacent to the stability boundary (and fewer on the other side of the disc), generating less complex and more structured dynamics than seen from a purely random matrix.
        
        \subsubsection{Local eigenvalue-outliers influencing destabilisation and non-trivial dynamics in sparse networks} \par
        If the connectivity matrix has structural (E-I) balance then the SZRS condition enforces a `tight' neuron-to-neuron input balance, and the network operates under balanced input conditions, i.e., dynamic balance, \cite{ipsen2020consequences,barral2016synaptic, landau2021macroscopic,ahmadian2021dynamical}. Note that the SZRS condition could be interpreted as a stricter condition than for the fully connected case because the condition effectively operates on all terms in the connectivity matrix. In this more strictly balanced case, destabilisation is accurately predicted by the eigenspectral radius, $\mathcal{R}$, crossing the imaginary axis. 

        Contrary to this, in the case of sparse \underline{inhibitory} dominated matrices obeying Dale's law, neuron-to-neuron balance is not satisfied, and a partial SZRS condition does not ensure this. As a result local eigenvalue-outliers escape the bulk, and destabilisation may not be predicted accurately by the radius, $\mathcal{R}$. An appropriate condition to constrain these eigenvalues and preserve inhibitory dominance (imbalance) remains an open problem. This inhibitory dominated regime due to structural E-I imbalance is thought to provide substrates for more complex dynamics to emerge, such as endogenous and pathological oscillations such as those seen in seizures \cite{barral2016synaptic,landau2018coherent,ipsen2020consequences}. 

        \subsection{Limitations and future research} \par
        The analysis presented in this paper examines the local stability of the network around the homogeneous equilibrium in the asymptotic limit. Therefore, this investigation yields insight into the transition to spontaneous activity, but the exact nature of such dynamics after the transition is not able to be analysed within this framework. Further, as discussed our analysis only strictly applies to homogeneous equilibria. Heterogeneous equilibrium solutions are dependent on the realisation of the synaptic connectivity matrix, so an additional dependency is introduced to the networked Jacobian. Specifically, the Jacobian Eq.~\ref{Equ:Jacobian} may no longer be statistically proportional to the connectivity matrix, $W$, as the matrix $\Phi'(\boldsymbol{x^*})$ can be heterogeneous and will influence the statistics of the networked Jacobian. The analysis framework we present is applicable to heterogeneous fixed points and different firing-rate functions as shown numerically in \cite{ostojic2014two,mastrogiuseppe2017intrinsically}. However, it is unclear how heterogeneous the matrix $\Phi'(\boldsymbol{x^*})$ has to be before it influences the statistics of the networked Jacobian and the results from random matrix theory no longer apply. 
        
        Another limitation of the network model used here is the assumption of instantaneous rise time in the post-synaptic potentials, i.e., there is no synaptic dynamics. A possible future extension is to incorporate synaptic dynamics, such as through the introduction of conductance-based synapses \cite{peterson2015homotopic}. Modelling synaptic dynamics is more realistic and could significantly change the dynamics through the additional feedback non-linearity. However, even though the network model used in this paper is not physiologically detailed, significant insight is generated about the neural system it describes, particularly in regards to the relationship between connectivity and dynamics.  
        
        In this paper, we examine randomly distributed connectivity weights that follow the product distribution of binomial and Gaussian random variables. However, connectivity weights in the cortex have been found to be log-normally distributed \cite{buzsaki2014log}. As this work is based on results from random matrix theory, which hold for any iid random variable \cite{tao2008random}, the results presented here should be extendable for any iid random variable, including log-normal random variables. 

        Currently, a condition to ensure the local eigenvalues do not escape the bulk disc for inhibitory/excitatory dominated (unbalanced) sparse networks does not exist. This is presently an open problem in random matrix theory, and future work would be to derive an appropriate condition to control these eigenvalues.

        The analysis presented here is performed for general ratios and distributions of excitation to inhibition, and probability of connection $\alpha$. For a local cortical network the typical ratio of excitatory to inhibitory connections is 4:1 \cite{braitenberg2013anatomy,kandel2000principles,shepherd2004synaptic}. Therefore, for structural (E-I) balance the strength of inhibition (i.e., the number of inhibitory synapses times their amplitude) must be four times that of excitation. A direct application of this analysis is to examine the connectivities of a local cortical network constructed with key statistics extracted from large connectomic data sets \cite{jones2009allen,paquola2021bigbrainwarp,shapson2021connectomic}. The statistics pertain to the ratio of inhibition to excitation, the mean and variance of the excitatory and inhibitory connections, and the average number of connections of a neuron to other neurons (sparsity). We further note that synaptic self-connections (autapses), are not very common \cite{bacci2006enhancement,bekkers2003synaptic}. However, we do not eliminate the self connection terms (diagonal terms) in the connectivity matrix, as the effect of removing these is negligible for large N. For finite sized networks, this should be a consideration particularly with analysis using connectomic data statistics.

        \subsection{Conclusion}
        In conclusion, network sparsity and Dale's law are two fundamental anatomical properties of local cortical networks in the brain. The respective impacts of these properties have been previously individually examined. This paper analyses their combined influences for structurally E-I balanced and unbalanced networks and demonstrates that balance and sparsity interact in ways that are counter-intuitive and have not previously been studied. We show that sparsity linearly scales the structural E-I balance of a connectivity matrix, and the eigenvalue outlier. However, in contrast to this, the variance of the connectivity matrix is a function of the nonlinear interaction between sparsity and the population means and variances. Therefore, the eigenvalue spectral disc radius also scales in this nonlinear fashion. Further, we find that the nonlinear interaction of sparsity with the population means and variances also influences the non-uniform eigenvalue spectral density. In this study, we also addressed the problem of local eigenvalue outliers and proved that these can be controlled for the unbalanced fully connected case and the balanced sparse case by deriving a new SZRS condition. We also provided some mathematical intuition behind why they cannot be controlled for the unbalanced sparse case, which remains an open problem. 
        In summary, our results indicate that there is a dynamical and non-linear interplay between network sparsity and all the E-I population statistics that is fundamental to regulating neural network dynamics. 
        
        The analysis presented here further develops the quantitative relationship between neural network architectures and neural dynamics. This relationship is of particular importance for both theoretical and experimental neuroscience as it pertains to the structure-function relationship found in local cortical networks. Our results are an important step towards developing analysis techniques that will be essential in understanding the impacts of larger scale network connectivity on brain function.

%         \\ \\
%         We provide an important step towards a fuller understanding of two of the key principles underlying network architecture in the brain, namely sparsity and Dale's law, and in developing analysis techniques for this trove of connectome data that is essential to understand and elucidate how connectome statistics influences neural network dynamics. 
% \\ \\

% \begin{acknowledgments}
\section*{Acknowledgements}
I.D.H. was supported by an Australian Government Research Training Program Scholarship provided by the Australian Commonwealth Government and the Graeme Clark Institute at the University of Melbourne. A.N.B. and H.M. were supported by the Australian Government through the Australian Research Council’s Discovery Projects funding scheme (Project DP220101166). A.P was supported by research fellowships from the Greame Clark Institute and St.Vincent's Hospital, Melbourne, Australia.  
% We wish to acknowledge the support of the author community in using
% REV\TeX{}, offering suggestions and encouragement, testing new versions,
% \dots.
% \end{acknowledgments}

\bibliography{apssamp}% Produces the bibliography via BibTeX.

\newpage
\appendix
\onecolumngrid

\section{\label{app:derivation1} Calculating the mean and variance of the entries of a sparse random matrix.}
        We calculate the expected value and variance of the entries in a sparse (un)balanced random matrix, $W$, as constructed in Eq.~\ref{equ:sWrankperturb} of the main text, with mean, $\mu$ and variance, $\sigma^{2}$ of the partially random component $W = AD + M$, and sparse component $S$ defined by the probability of connection $\alpha$. The expected value of the entries in $W$ takes the form
            \begin{align}
            \EX(w_{ij}) &= \frac{1}{N^2} \sum_{i=1}^{N}\sum_{j=1}^{N} w_{ij} \nonumber \\
                      &= \frac{1}{N^2} \sum_{i=1}^{N}\sum_{j=1}^{(1-\alpha)N} 0 + \frac{1}{N^2} \sum_{i=1}^{N}\sum_{j=1}^{\alpha N} \hat{w}_{ij} \nonumber \\
                      &= \alpha \mu.
            \end{align}
            The variance takes the form
            \begin{eqnarray}
            \VAR(w_{ij}) &&= \frac{1}{N^2} \sum_{i=1}^{N}\sum_{j=1}^{N} w_{ij}^{2} - \left(\frac{1}{N^2} \sum_{i=1}^{N}\sum_{j=1}^{N} w_{ij}\right)^{2} \nonumber \\
                      &&= \frac{1}{N^2} \sum_{i=1}^{N}\sum_{j=1}^{(1-\alpha)N} 0^2 + \frac{1}{N^2} \sum_{i=1}^{N}\sum_{j=1}^{\alpha N} \hat{w}_{ij}^{2} - \alpha^{2} \mu^{2} \nonumber\\
                      &&= \frac{1}{N^2} \sum_{i=1}^{N}\sum_{j=1}^{\alpha N} \hat{w}_{ij}^{2} - \alpha^{2} \mu^{2} \nonumber \\
                      &&= \frac{1}{M} \sum_{k=1}^{\alpha M} m_{\mathrm{k}}^{2} - \alpha^{2} \mu^{2} \label{Aequ:subs1}\\
                      &&= \alpha \frac{1}{\alpha M} \sum_{k=1}^{\alpha M} m_{\mathrm{k}}^{2} - \alpha^{2} \mu^{2} \nonumber \\
                      &&= \alpha \left(\mu^{2} + \sigma^{2}\right) - \alpha^{2} \mu^{2} \label{Aequ:moment1}
            \end{eqnarray}
            where we set $M=N^2,m_{\mathrm{k}}=\hat{w}_{i,j}$ in Eq.~\ref{Aequ:subs1} and note that $\frac{1}{\alpha M} \sum_{k=1}^{\alpha M} m_{\mathrm{k}}^{2}$ is the second non-central moment of $\mathcal{N}(\mu,\sigma)$ Eq.~\ref{Aequ:moment1}. Hence, the expressions for $\EX(w_{ij})$ and $\VAR(w_{ij})$ are
            \begin{equation}
                \EX(w_{ij}) = \alpha \mu,
                \quad
                \VAR(w_{ij}) = \alpha \left(\mu^{2} + \sigma^{2}\right) - \alpha^{2} \mu^{2} 
            \end{equation}

\section{\label{app:derivation2} Calculating the mean and variance of the entries of a sparse random matrix obeying Dale's law.} 
    We commence by separately calculating the means of the excitatory and inhibitory weights in $W$ constructed as in Eq.~\ref{equ:sWrankperturb}.
        \begin{eqnarray}
            \EX(w_{ij})_{\mathrm{e}} &&= \frac{1}{fN^2} \sum_{i=1}^{N}\sum_{j=1}^{fN} w_{ij} \nonumber \\
                    &&= \frac{1}{fN^2} \sum_{i=1}^{N}\sum_{j=1}^{f(1-\alpha)N} 0 + \frac{1}{fN^2} \sum_{i=1}^{N}\sum_{j=1}^{f\alpha N} w_{ij} \nonumber\\
                    &&= \alpha \mu_{\mathrm{e}} \\
            \EX(w_{ij})_{\mathrm{i}} &&= \frac{1}{(1-f)N^2} \sum_{i=1}^{N}\sum_{j=1}^{(1-f)N} w_{ij} \nonumber\\
                    &&= \frac{1}{(1-f)N^2} \sum_{i=1}^{N}\sum_{j=1}^{(1-f)(1-\alpha)N} 0 \nonumber\\
                    && \quad + \frac{1}{(1-f)N^2} \sum_{i=1}^{N}\sum_{j=1}^{(1-f)\alpha N} w_{ij} \nonumber\\
                    &&= \alpha \mu_{\mathrm{i}}
        \end{eqnarray}
    The mean of the entries in the matrix $W$ is simply the weighted sum of the means of the excitatory and inhibitory weights. This is expressed as follows,
        \begin{equation}
            \EX(W) = f  \mu_{\mathrm{se}} + (1-f) \mu_{\mathrm{si}}
        \end{equation}
    where $\mu_{\mathrm{sk}} = \alpha \mu_{\mathrm{k}}$ are the means of the excitatory and inhibitory weights $k=\mathrm{e}, \mathrm{i}$.

    We now calculate the variances for each neural population and we substitute the second non-central moment of the population distribution to obtain separate variance expressions for the excitatory and inhibitory weights.
        \begin{eqnarray}
        \VAR(w_{ij})_{\mathrm{e}} &&= \frac{1}{fN^2} \sum_{i=1}^{N}\sum_{j=1}^{fN} w_{ij}^{2} - \left(\frac{1}{fN^2} \sum_{i=1}^{N}\sum_{j=1}^{fN} w_{ij}\right)^{2} \nonumber\\
                    &&= \frac{1}{fN^2} \sum_{i=1}^{N}\sum_{j=1}^{f(1-\alpha)N} 0^2 + \frac{1}{fN^2} \sum_{i=1}^{N}\sum_{j=1}^{f\alpha N} \hat{w}_{ij}^{2} \nonumber \\
                    && \quad - \alpha^{2} \mu_{\mathrm{e}}^{2} \nonumber\\
                    &&= \frac{1}{fM} \sum_{k=1}^{f\alpha M} m_{\mathrm{k}}^{2} - \alpha^{2} \mu_{\mathrm{e}}^{2} \nonumber\\
                    &&= \alpha \frac{1}{f\alpha M} \sum_{k=1}^{f \alpha M} m_{\mathrm{k}}^{2} - \alpha^{2} \mu_{\mathrm{e}}^{2} \nonumber\\
                    &&= \alpha \left(\mu_{\mathrm{e}}^{2} + \sigma_{\mathrm{e}}^{2}\right) - \alpha^{2} \mu_{\mathrm{e}}^{2}
        \end{eqnarray}
        \begin{eqnarray}
        \VAR(w_{ij})_{\mathrm{i}} &&= \frac{1}{(1-f)N^2} \sum_{i=1}^{N}\sum_{j=1}^{(1-f)N} w_{ij}^{2} \nonumber \\
                    && \quad - \left(\frac{1}{(1-f)N^2} \sum_{i=1}^{N}\sum_{j=1}^{(1-f)N} w_{ij}\right)^{2} \nonumber\\
                    &&= \frac{1}{(1-f)N^2} \sum_{i=1}^{N}\sum_{j=1}^{(1-f)(1-\alpha)N} 0^2 \nonumber \\
                    && \quad + \frac{1}{(1-f)N^2} \sum_{i=1}^{N}\sum_{j=1}^{(1-f)\alpha N} \hat{w}_{ij}^{2} - \alpha^{2} \mu_{\mathrm{i}}^{2} \nonumber\\
                    &&= \frac{1}{(1-f)M} \sum_{k=1}^{(1-f)\alpha M} m_{\mathrm{k}}^{2} - \alpha^{2} \mu_{\mathrm{i}}^{2} \nonumber \\
                    &&= \alpha \frac{1}{(1-f)\alpha M} \sum_{k=1}^{(1-f) \alpha M} m_{\mathrm{k}}^{2} - \alpha^{2} \mu_{\mathrm{i}}^{2} \nonumber\\
                    &&= \alpha \left(\mu_{\mathrm{i}}^{2} + \sigma_{\mathrm{i}}^{2}\right) - \alpha^{2} \mu_{\mathrm{i}}^{2}
        \end{eqnarray}
    
    The variance of the entries in the matrix $W$ is the weighted sum of the variance of the excitatory and inhibitory weights. We express this as
    \begin{equation}
        \VAR(W) = f\sigma_{\mathrm{se}}^{2} + (1-f) \sigma_{\mathrm{si}}^{2},
    \end{equation}
    where $\sigma_{\mathrm{sk}}^{2} = \alpha (1-\alpha)\mu_{\mathrm{k}}^{2} + \alpha \sigma_{\mathrm{k}}^{2} $ are the variances of the excitatory and inhibitory weights $k=\mathrm{e}, \mathrm{i}$.

\section{\label{app:additionalfigures} Additional Figures}
            \begin{figure}[ht]
                \centering
                    \includegraphics[width=0.5\columnwidth]{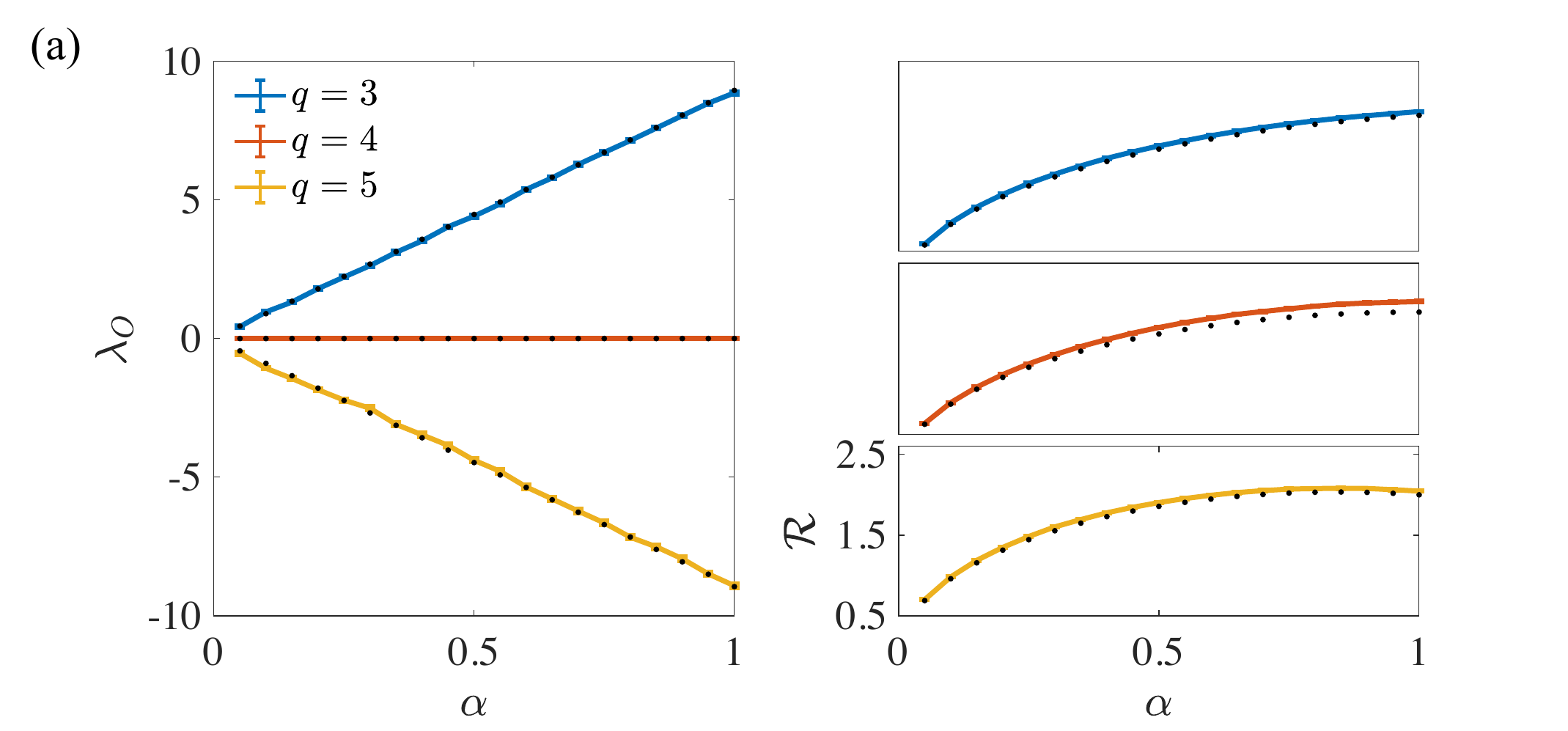}%
                \quad
                    \includegraphics[width=0.5\columnwidth]{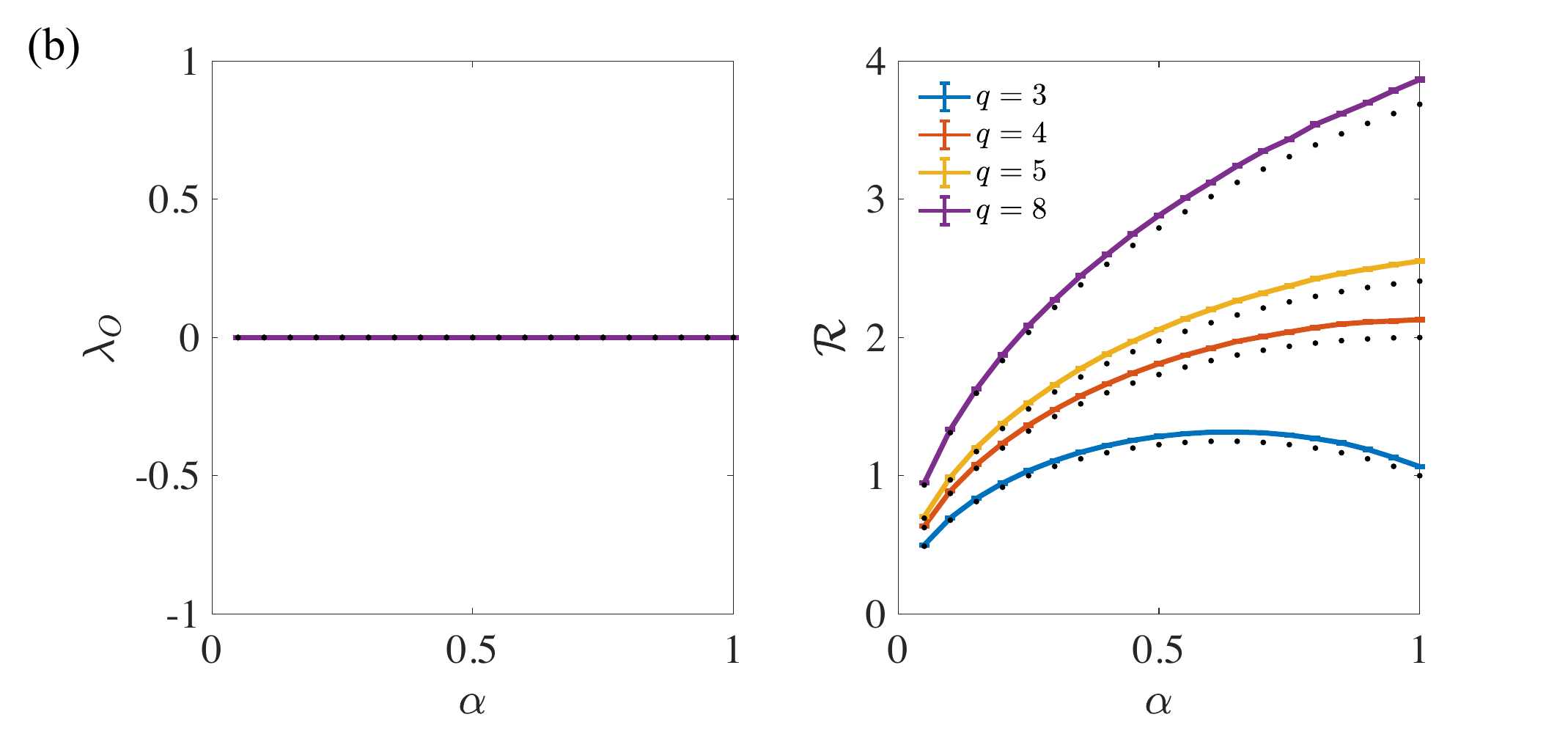}%
                 \quad
                    \includegraphics[width=0.5\columnwidth]{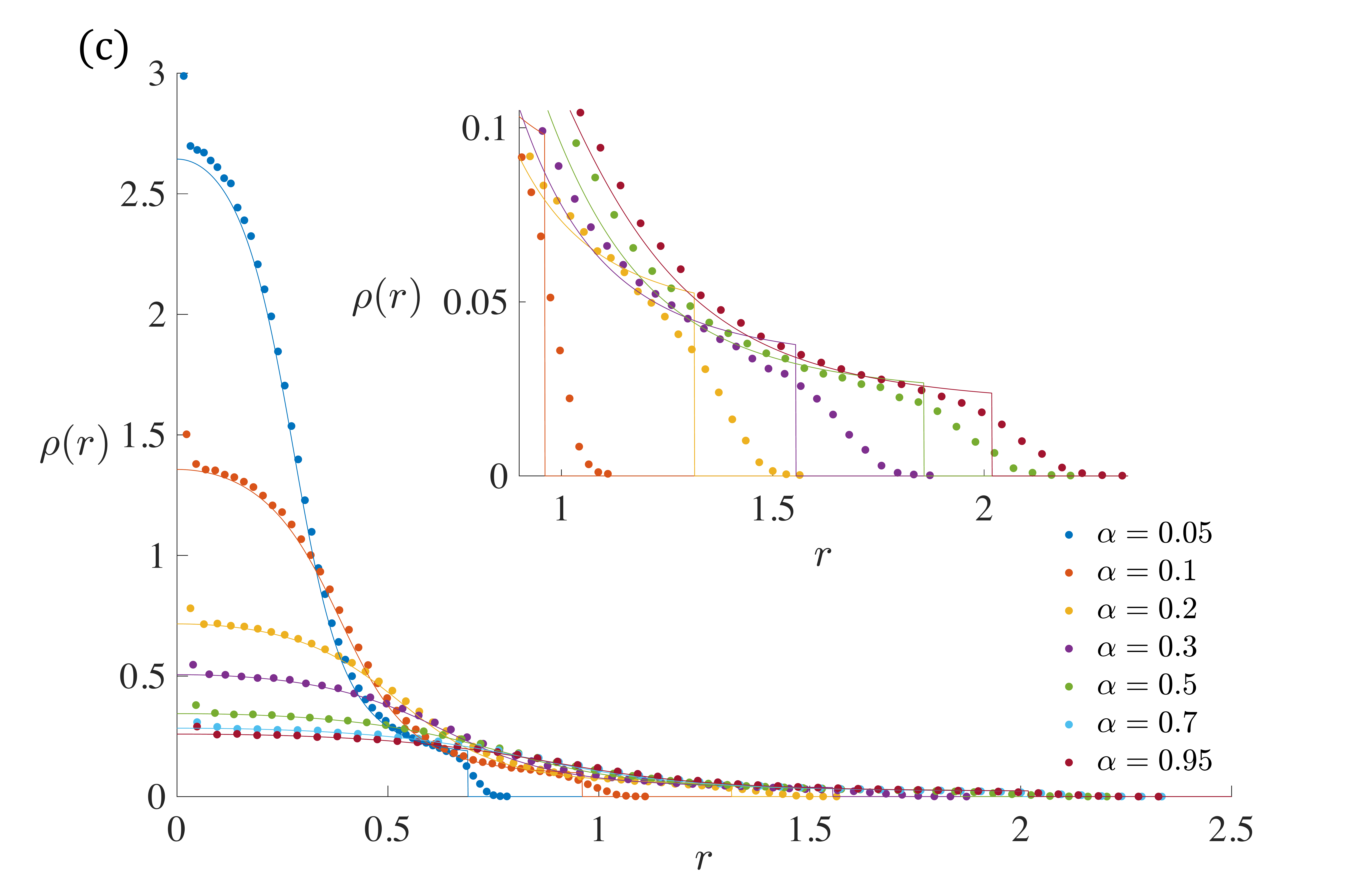}%
            \caption{This extends the results in Figure~\ref{fig:twopopfigOR} to the structurally E-I unbalanced case. (a-b) The eigenvalue outlier location (left panel) and radius of the eigenspectral-disc (right panel) for the matrix defined in Eq.~\ref{equ:sWrankperturb}, as a function of the sparsity probability $\alpha$ with $N = 2000, f=0.8$ and $q=3,4,5,8$. The theoretical outlier location and radius were calculated using definitions in Eq.~\ref{equ:lambdao2} and \ref{equ:rad2a}, respectively. The black dots represent the numerical eigenvalue outlier and radius averaged over 100 realisations.
            (a) Plots the case where the inhibitory mean is varied $\mu_{\mathrm{i}} = -\frac{q}{\sqrt{N}}$, and all other network parameters are held constant $\mu_{\mathrm{e}} = \frac{1}{\sqrt{N}},\sigma_{\mathrm{e}} = \sigma_{\mathrm{i}} = \frac{1}{\sqrt{N}}$. (b) Shows the case where the inhibitory variance is changed $\sigma_{\mathrm{i}} = \frac{q}{\sqrt{N}}$, and all other network parameters held constant $\mu_{\mathrm{e}} = \frac{1}{\sqrt{N}}, \mu_{\mathrm{i}} = -\frac{4}{\sqrt{N}}$, $\sigma_{\mathrm{e}} = \frac{1}{\sqrt{N}}$. The case where the eigenvalue outlier is zero indicates a structurally E-I balanced network: orange in (a), and all colours in (b). (c) Non-uniform spectral density curves as a function of normalised disc radius and sparsity parameter $\alpha$ with analytical expression as solid lines and numerical simulations as points. Parameters of $W$ are $N = 2000$, $\mu_{\mathrm{e}} = \frac{1}{\sqrt{N}}, \mu_{\mathrm{i}} = -\frac{5}{\sqrt{N}}$, $\sigma_{\mathrm{e}} = \frac{1}{\sqrt{N}}, \sigma_{\mathrm{i}} = \frac{4}{\sqrt{N}}$, and $f=0.8$.}
            \label{fig:twopopmeansvarOR}
            \end{figure}

            % \begin{figure}[ht]
            % \centering
            %   \includegraphics[width = 0.75\columnwidth]{N1000homotopycont_r250_g5_4.pdf}%
            %     \caption{ Numerical homotopy illustrating the average proportion of eigenvalue crossings as a function of the continuous deformation from $\kappa=0$ to $\kappa \rightarrow \sigma$.  The proportion of eigenvalue crossings is an average calculated over $r=250$ realisations, and the standard error is included as bars on each point. The homotopy parameter $\kappa$ controls the separation of the means of the two populations, $\mu_{\mathrm{e}} = \kappa$, $\mu_{\mathrm{i}} = -5\kappa$, and the population variances are $\sigma_{\mathrm{e}}^{2} = 1/N, \sigma_{\mathrm{i}}^{2} = 4/N$. The matrix has dimension $N=1000$ and the proportion of the excitatory population is $f=0.8$. Each sub-figure shows the homotopy for different values of $\alpha$ where alpha ranges from $0.1-1$.}
            %     \label{fig:N2000homotopy_unbalanced}
            % \end{figure}

%\nocite{*}

\end{document}